 \documentclass[preprint2,longabstract]{aastex}

\slugcomment{Submitted to the Astrophysical Journal}

\shorttitle{Electron Excitation and Nature of Molecular Gas in Cluster Central Ellipticals}
\shortauthors{Lim et al.}

\usepackage{multirow}
\usepackage{color}
\usepackage{soul}
\usepackage[normalem]{ulem}
\usepackage{breqn}

\begin{document}


\title{Role of Electron Excitation and Nature of Molecular Gas \\ in Cluster Central Elliptical Galaxies}


\author{Jeremy Lim\altaffilmark{1}}
\affil{Department of Physics, The University of Hong Kong, Pokfulam Road, Hong Kong \\
\& \\
Laboratory for Space Research, Faculty of Science, The University of Hong Kong, Pokfulam Road, Hong Kong}\email{jjlim@hku.hk}

\author{Dinh-V-Trung}
\affil{Institute of Physics, Vietnam Academy of Science and Technology, 10 DaoTan, ThuLe, BaDinh, Hanoi, Vietnam}
\email{dvtrung@iop.vast.ac.vn}

\author{Jan Vrtilek}
\affil{Harvard-Smithsonian Center for Astrophysics, 60 Garden St., Cambridge, MA 02138, USA}
\email{jvrtilek@cfa.harvard.edu}

\author{Laurence P. David}
\affil{Harvard-Smithsonian Center for Astrophysics, 60 Garden St., Cambridge, MA 02138, USA}
\email{ldavid@cfa.harvard.edu}

\and  

\author{William Forman}
\affil{Harvard-Smithsonian Center for Astrophysics, 60 Garden St., Cambridge, MA 02138, USA}
\email{wforman@cfa.harvard.edu}


\altaffiltext{1}{Adjunct Research Fellow, Institute of Astronomy \& Astrophysics, Academia Sinica, Taipei 10617, Taiwan}


\clearpage
\newpage

\begin{abstract}
We present observations in CO(3-2) that, combined with previous observations in CO(2-1), constrain the physical properties of the filamentary molecular gas in the central $\sim$6.5\,kpc of \object{NGC 1275}, the central giant elliptical galaxy of the Perseus cluster.  We find this molecular gas to have a temperature $\gtrsim 20$\,K and a density $\sim$$10^2$--$10^4 {\rm \ cm^{-3}}$, typically warmer and denser than the bulk of Giant Molecular Clouds (GMCs) in the Galaxy.  Bathed in the harsh radiation and particle field of the surrounding intracluster X-ray gas, the molecular gas likely has a much higher ionization fraction than that of GMCs.  For an ionization fraction of $\sim$$10^{-4}$, similar to that of Galactic diffuse ($\lesssim 250 {\rm \ cm^{-3}}$) partially-molecular clouds that emit in HCN(1-0) and HCO$^+$(1-0), we show that the same gas traced in CO can produce the previously reported emissions in HCN(3-2), HCO$^+$(3-2), and CN(2-1) from \object{NGC 1275}; the dominant source of excitation for all the latter molecules is collisions with electrons.  To prevent collapse, as evidenced by the lack of star formation in the molecular filaments, they must consist of thin strands that have cross-sectional radii $\lesssim$0.2--2\,pc if supported solely by thermal gas pressure; larger radii are permissible if turbulence or poloidal magnetic fields provide additional pressure support.  We point out that the conditions required to relate CO luminosities to molecular gas masses in our Galaxy are unlikely to apply in cluster central elliptical galaxies.  Rather than being virialized structures analogous to GMCs, we propose that the molecular gas in \object{NGC 1275} comprises pressure-confined structures created by turbulent flows.

\end{abstract}





\section{Introduction}\label{Introduction}
The central giant elliptical galaxies of cool-core clusters\footnote{A designation introduced by \citet{mol01} for galaxy clusters that exhibit a temperature decrement in their intracluster X-ray emitting gas at the cluster core.  The X-ray gas permeating such clusters exhibits a surface brightness that is strongly centrally peaked, indicating prodigious radiative loss at the cluster core.  The resulting loss in pressure support was predicted to result in an inflow of cooling intracluster gas, referred to as an X-ray cooling flow \citep{cow77,fab77}.  \citet{mol01} found, as demonstrated more robustly by \citet{dav01} and \citet{pet03}, that the mass-deposition rate from any such flow is much lower than had previously been inferred.  Today, it is widely recognized that X-ray cooling flows can be mitigated, if not entirely quenched, by powerful jets from the supermassive black hole in the cluster central elliptical galaxy \citep[e.g., review by][]{fab12}.} preferentially emit atomic and ionic lines at optical wavelengths \citep[e.g.,][]{hec89,cra99,cav08,don10}.  Their optical spectra have long been known to be peculiar, exhibiting both low and high excitation lines simultaneously from neutral atoms and ions.  
A number of these galaxies also exhibit molecular lines as detected in the ro-vibrational lines of H$_2$ \citep[e.g.,][]{edg01} and/or CO \citep[e.g.,][]{edg02,sal03}.  Their CO luminosities can rival that of ultraluminous infrared galaxies, placing cluster central elliptical galaxies among the most CO-lumimous galaxies in the Local Universe.

Of these galaxies, the best studied example is \object{NGC 1275}, the central giant elliptical galaxy of the Perseus cluster, the X-ray-brightest cluster in the sky.  This galaxy first gained prominence through the discovery by \citet{min55} of a set of optical emission lines associated with \object{NGC 1275} together with another set associated with a foreground galaxy redshifted by $\sim$$3000 {\rm \ km \ s^{-1}}$ and located just north of the center of \object{NGC 1275}, a frequently cited example to bolster early claims that radio galaxies are triggered by violent interactions.  The full grandeur of the optical line-emitting nebula associated with \object{NGC 1275} was not revealed until the photograph taken by \citet{lyn70}, with subsequent improvements in clarity provided by the CCD images taken by \citet{con01} with the WIYN 3.5 m telescope and \citet{fab08} with the Hubble Space Telescope.  Figure\,\ref{NGC 1275} shows a combined multi-color and H$\alpha$+[NII] image of \object{NGC 1275} taken by R. Jay Gabany with the 0.5 m telescope of the Blackbird Observatory.  As can be seen, the nebula has a complex morphology, comprising numerous filaments that are exactly or nearly radial, but also some that are approximately tangential or clearly curved (shaped like, or part of, a ``U").  The atomic and ionic components detected in optical, as well as infrared \citep{joh07}, emission lines (perhaps at $\sim$$10^4$\,K) have since been shown to be accompanied by ionized gas (at perhaps $\sim$5--10\,MK) detected in X-rays \citep{fab06,fab11}, ``hot" (at perhaps $\sim$2000\,K) molecular gas detected in the near-IR ro-vibrational lines of H$_2$ \citep{hat05,lim12}, ``warm" (at perhaps $\sim$300\,K) molecular gas detected in the mid-IR rotational lines of H$_2$ \citep{joh07}, and ``cool" (at perhaps in the range $\sim$10--200\,K) molecular gas detected in CO \citep{laz89,mir89,bri98,sal06,lim08,sal08a,sal08b,hol09,sal11}.

The manner by which the different line-emitting gas components in cluster central elliptical galaxies are excited remains contentious \citep[for a brief review, see][]{lim12}.  As a consequence, the temperature and density, and therefore also mass, of the individual gas components remain uncertain.  In \object{NGC 1275}, newly-formed massive stars are clearly not responsible, in general, for exciting its line-emitting nebula: as can be seen in Figure\,\ref{NGC 1275}, the relatively young stars traced in blue light are clearly displaced from, if at all associated with, the H$\alpha$ nebula shown in red \citep[see also][]{fab08,can10,can16,yul15}.  \citet{can14} showed that even star clusters having ages of less than $\sim$10\,Myr are generally not cospatial with the line-emitting nebula.  Because luminous line-emitting nebulae in cluster central elliptical galaxies are preferentially if not exclusively associated with those in cool-core clusters, these nebulae provide the best --- albeit circumstantial --- evidence for cooling of the intracluster medium (ICM).  The uncertainties in their physical properties, however, frustrate their use for quantifying physical processes at work in the cluster core (e.g., gas-cooling rate) as well as in the cluster central elliptical galaxy (e.g., gas reservoir and potential for star formation).

To explain the unusual optical to infrared spectra of cluster central elliptical galaxies, \citet{fer08,fer09} argue that energetic particles must play an important role in setting the physical and excitation state of their interstellar medium (ISM).  Such particles can ionize as well as dissociate molecular hydrogen (H$_2$), and ionize atomic hydrogen (H), to produce a gas mixture comprising H$_2$, H$_2^+$, H, and H$^{+}$ (protons), along with the ionized electrons.  If the highly-energetic particles comprise mainly protons, as in cosmic rays, then collisions between these particles and atomic or molecular gas first release a primary population of energetic electrons, which then collide repeatedly with mainly H and/or H$_2$ to create a secondary population of less-energetic, albeit suprathermal, electrons.  In such a gas mixture, the composition (i.e., fraction in ions, atoms, and molecules) and temperature of which depends on the density of the energetic particles, atoms and molecules (both neutral and ions) can be excited primarily by collisions with the ionized electrons rather than through radiation or collisions with H and/or H$_2$.  Furthermore, the abundance of certain ionic species can be radically altered through charge exchange with H, thus suppressing some ionic lines having relatively low ionization potentials but not affecting other ionic lines having higher ionization potentials.  In the case of \object{NGC 1275}, \citet{fer09} showed that a judicious combination of gas mixtures having different compositions and occupying a broad range of temperatures can purportedly explain its unusual optical and infrared spectra.  By contrast with normal star-forming galaxies where the H$\alpha$ line is produced primarily through recombination in highly-ionized gas at a temperature of $\sim$$10^4$\,K (e.g., HII regions, supernova remnants, and diffuse photodissociated gas), in \object{NGC 1275} this line may be produced predominantly if not entirely by cascades following electron-impact excitation of H in primarily neutral gas spanning a broad range of densities and temperatures \citep[$\sim$$10^2$--$10^4$\,K according to the model of][]{fer09}.  Indeed, many of the optical and infrared emission-lines from trace atomic and ionic species may originate, at least in part, from regions where H is mostly neutral and at relatively low temperatures.  Similarly, H$_2$ lines at mid-/near-infrared wavelengths, rather than being excited through mutual collisions (as well as collisions with H) at temperatures of $\sim$300/2000\,K, may instead be excited by electrons in gas spanning a broad range of densities and temperatures \citep[$\sim$10--1000\,K in the model of][]{fer09} where collisions with H$_2$ or H may not be able to overcome the energy threshold for exciting these lines.  Recently, \citet{can16} extended these computations to predict the emissivities of bright atomic lines at far-infrared wavelengths and a number of molecular lines (from CO, HCN, HCO$^+$) at sub-millimeter wavelengths.

Although \citet{fer09} only speculated on possible sources for the hypothesized highly-energetic particles, \citet{fab11} has advocated that, in at least the case of \object{NGC 1275}, the surrounding X-ray gas which penetrates into the line-emitting filaments \citep[which is believed to be threaded by strong magnetic fields;][]{fab08} constitutes the source of energetic particles \citep[see also discussion in][]{lim12,fab12}.  As evidence, \citet{fab11} point out that the surface radiative flux from the filaments is close to the thermal energy flux impacting on them from the surrounding X-ray gas if this gas penetrates into the filaments at an efficiency of about 10\%.  In this scenario, the X-ray component of the filaments at perhaps $\sim$5--10\,MK \citep{fab06,fab11} may constitute the product of interactions between the infiltrating X-ray gas and the cooler ISM gas components. 

A gas component of broad interest in the ISM of any galaxy is cool molecular gas, the reservoir for star formation.  Because collisional excitation of H$_2$ requires densities ($\gtrsim$$10^{5-6} {\rm \ cm^{-3}}$) and temperatures ($\gtrsim 10^3 {\rm \ K}$) much higher than are characteristic of the bulk molecular gas in star-forming regions, trace molecules that can be excited by collisions with H$_2$ even at relatively low temperatures ($\gtrsim$10\,K) are used to trace cool molecular (i.e., predominantly H$_2$) gas.
The most abundant trace molecule, CO, can be strongly excited at the relatively low densities ($\gtrsim 10^2 {\rm \ cm^{-3}}$) and temperatures of Galactic molecular clouds where stars form, and hence is widely used to infer the gas reservoir for star formation.  Stars in our Galaxy, however, form in dense condensations (which, in the field of star formation, is referred to as dense cores) that occupy just a small fraction in volume and comprise just a small fraction in mass of molecular clouds.  Such condensations are usually traced in molecules such as HCN, HCO$^+$, or CN, all of which require higher gas densities ($\gtrsim$$10^{5-6} {\rm \ cm^{-3}}$) to be strongly excited.  Emission from all these molecules have been detected from \object{NGC 1275} \citep{sal08b,bay11}, suggesting the presence of relatively dense molecular gas (located presumably within the bulk molecular gas traced by CO) in this galaxy.  Yet, newly-formed massive stars are generally not found in the molecular gas traced by CO \citep{hol09}, which when mapped at a sufficiently high angular resolution has been found to be cospatial with the optical line-emitting filaments \citep{sal06,lim08,sal08a,sal08b,hol09}.  This departure from the situation normally seen in star-forming galaxies, together with the argument that energetic electrons play an important of not dominant role in exciting both the optical line-emitting gas as well as the infrared H$_2$ lines, motivate us to examine explicitly whether the same electrons also play a role in exciting trace molecules in cluster central elliptical galaxies.  As we will show, although CO is excited primarily by collisions with H$_2$ as in Galactic molecular clouds, HCN, HCO$^+$, and CN can be excited primarily by collisions with electrons rather than H$_2$, and therefore need not necessarily trace dense molecular gas.  Indeed, both HCN(1-0) and HCO$^+$(1-0) have been detected from diffuse partially-molecular clouds in our Galaxy, where the H$_2$ gas density is $\lesssim$\,$250 {\rm \ cm^{-3}}$ but the ionization fraction $\sim$$10^{-4}$ \citep[e.g.,][and references therein]{lis12}\footnote{The ionization fraction of these clouds is determined by the elemental abundance of carbon, which is nearly fully once-ionized (i.e., in the form of C\,II), with a contribution from cosmic-ray ionization of hydrogen.}.  In such clouds, the molecules that give rise to these transitions are believed to be excited primarily through collisions with electrons.

In addition, we present observations of \object{NGC 1275} in CO(3-2) that, when combined with our previously published observations in CO(2-1), allow us to determine the ratio in CO(3-2) to CO(2-1) on a spatial scale of $\sim$1\,kpc.  By comparing these measurements with a simple radiative transfer model, we constrain the physical properties of the CO-emitting molecular gas --- rather than using an uncertain conversion between CO luminosity and H$_2$-gas mass --- towards the inner regions of \object{NGC 1275}.  Including also collisional excitation by electrons in addition to H$_2$ molecules, we show explicitly that HCN(3-2) --- and, by implication, also HCO$^+$(3-2) and CN(2-1) --- can arise from the same molecular gas as that traced by CO provided the ionization fraction of this gas is $\sim$$10^{-4}$ or higher.  Thus, the detection of molecules that are conventionally regarded as tracers of dense gas akin to star-forming condensations in Galactic molecular clouds may not always be a good predictor for star formation in cluster central elliptical galaxies; the molecular (and atomic) gas in these galaxies are bathed in copious ionizing radiation from the surrounding X-ray gas, which furthermore can infiltrate and excite (through the production of secondary suprathermal electrons) as well as heat the molecular (and atomic) gas.  
Readers interested in how the CO(3-2) observations were conducted and the data reduced should proceed to $\S2$.  The results in CO(2-1) have previously been presented by \citet{hol09}, which includes the data taken by \citet{lim08}.  
The results for CO(3-2), and the measured ratio in brightness temperature between CO(3-2) and CO(2-1), are given in $\S3$.  In $\S4$, we consider the different ways in which CO, HCN, HCO$^+$, and CN can be excited.  In $\S5$, we examine the implications of our measurements in CO(3-2) and CO(2-1) for the physical conditions of the molecular gas traced by these lines in \object{NGC 1275}.  We also examine the required ionized fraction of this gas if it also produces the observed emission in HCN(3-2), HCO$^+$(3-2), and CN(2-1).  In $\S6$, we review the two diametrically opposing views for the nature of molecular clouds as traced by CO in our Galaxy, and how these views inform the physical basis for a relationship between CO luminosity and H$_2$-gas mass of these clouds.  Our intentions are to make clear and explicit both the arguments and assumptions implicit in the widely-used conversion between CO luminosity and H$_2$-gas mass.  We examine whether this conversion, as derived from observations of molecular clouds in our Galaxy, can be applied to \object{NGC 1275} as well as other cluster central elliptical galaxies.  
Finally, in $\S7$, we reexamine the stability of the molecular component in the filamentary nebula of \object{NGC 1275} based on its physical properties as inferred from our measurements rather than through the CO luminosity to H$_2$-gas mass conversion.  A concise summary of our results and conclusions can be found in $\S8$.  Throughout this manuscript, we assume a distance to \object{NGC 1275} of 74~Mpc, so that $1\arcsec = 360 {\rm \ pc}$.  All velocities quoted are referenced with respect to the heliocentric velocity of \object{NGC 1275}; in the optical definition, the heliocentric velocity is $cz = 5264 {\rm \ km \ s^{-1}}$ \citep{huc99}.

\section{Observations and Data Reduction}\label{Observations}
We observed \object{NGC 1275} in CO(3-2) with the SubMillimeter Array (SMA) in its compact configuration on 2006 December 16.  At the time of our observation, six of the eight antennas in the array were operational, providing 15 baselines ranging in projected separations from $\sim$7\,m to $\sim$64\,m.  The dual-sideband receivers operating in the 345~GHz (0.8\,mm) band were tuned to a central observing frequency of 329.798\,GHz in the lower sideband and 339.798\,GHz in the upper sideband, thus placing the CO(3-2) line in the upper sideband.  The correlator was configured to cover 24 spectral windows (chunks in the SMA nomenclature) with 128 channels in each chunk.  This configuration provided a spectral resolution of 0.812\,MHz ($0.104 {\rm \ km \ s^{-1}}$) over a total bandwidth, with some overlap at the edges of adjacent chucks, of 1.98\,GHz ($1719 {\rm \ km \ s^{-1}}$).  At the frequency of the CO(3-2) line, the primary beam of the telescope as measured at full-width at half-maximum (FWHM) is 36.2\arcsec, and so our observations span a central region of diameter 13.0\,kpc.

We used both 0418+380 (3C~111) and Saturn for bandpass calibration.  For absolute flux calibration, and indeed to generate complex gain solutions for \object{NGC 1275}, we used an observationally-derived model for its nuclear continuum source, which is continually monitored at the SMA.  At the time of our observation, the nuclear continuum source had a flux density of 2.4\,Jy at 0.8\,mm.

\section{Results and Analyses}\label{Results}

\subsection{CO(3-2) spatial-kinematic structure}\label{structure}
Using natural weighting, the synthesized beam of the CO(3-2) maps has a size at full-width half-maximum (FWHM) of $2\farcs40 \times 2\farcs19$ (position angle of $9\degr.1$), somewhat smaller than the angular resolution of the CO(2-1) maps published in \citet{hol09} of $3\farcs43 \times 2\farcs75$ (position angle of $19\degr.9$).  To compute the ratio in brightness temperatures between these two lines, all the CO(3-2) maps shown in this manuscript have been convolved to the same synthesized beam as the CO(2-1) maps.  The root-mean-square (rms) noise level of the resulting CO(3-2) channel maps in each $20 {\rm \ km \ s^{-1}}$ channel is $\sim$21\,mJy, compared with that in the CO(2-1) channel maps in each $20 {\rm \ km \ s^{-1}}$ channel of $\sim$15\,mJy.  Figure\,\ref{CO3-2_channelmaps_mom0_col} shows contour plots of the channel maps in CO(3-2) overlaid on a color map of the integrated line intensity (see description below for how we generated the latter map).  The integrated-intensity map has been corrected for the primary beam response of the telescope, although not the channel maps so as to preserve the clarity of these maps.  Figure\,\ref{CO3-2_mom0_mom1} shows the same integrated-intensity (moment 0) map in contours overlaid on an intensity-weighted mean-velocity (moment 1) map in CO(3-2).  To make these moment maps, we first smoothed each channel map in velocity (Hanning function with a FWHM of $100 {\rm \ km \ s^{-1}}$) and space (Gaussian function with a FWHM of 2\farcs4), and then retained only those pixels that in given channel map exceed a selected threshold of 26\,mJy in the corresponding smoothed map.  To preserve clarity, the channel maps were not corrected for the primary beam response before computing the intensity-weighted mean-velocity map (any effect on the latter map would be small and restricted to near the edge of the primary beam).  

To provide a simple comparison with CO(2-1), in Figure\,\ref{CO3-2_mom0_CO2-1_mom0} we overlay in contours the CO(3-2) integrated-intensity map shown in the previous figures on a color image of the CO(2-1) integrated-intensity map reported by \citet{hol09}.  The different features are labeled according to the nomenclature introduced in \citet{lim08} and \citet{hol09}.  As can be seen, all the structures that we detect in CO(3-2) were previously detected in CO(2-1).  Within the primary beam of the CO(3-2) observation, the only structures detected in CO(2-1) not detected in CO(3-2) lie close to the edge of the primary beam, where the sensitivity drops off steeply.  


\subsection{Ratio CO(3-2) to CO(2-1)}\label{line ratio}
A central tenet for deriving physical parameters from line ratios is that the lines originate from the same parcel(s) of gas.
As a simple check of whether the CO(3-2) and CO(2-1) emissions likely originate from the same regions, we compare whether these two lines have the same spectral profile along a given line of sight.  In Figure\,\ref{line_profiles}, we plot the spectral profiles in both CO(3-2) and CO(2-1) at selected locations marked by the symbol "$+$" (nucleus) and the four symbols "$\times$" in Figure\,\ref{CO3-2_mom0_CO2-1_mom0}.  Their positions, measured with respect to that of the nuclear radio continuum source (at $0\arcsec$,\,$0\arcsec$), are listed in Table\,\ref{line parameters}, and correspond to just east of the center ($1\farcs7$,\,$0\arcsec$) towards the eastern end of the Inner filament, the approximate midpoint of the Inner filament ($-3\farcs5$,\,$0\arcsec$), the eastern (inner) end of the Western filament ($-9\farcs5$,\,$1\farcs4$), and a region close to both the midpoint of the Western filament as well as the compact feature W2 ($-13\farcs0$,\,$1\farcs4$).  Note that the locations corresponding to the center of the galaxy and just east of the center are separated by only approximately half the synthesized beam, and hence the spectral profile at each location contains a contribution from the adjacent location.  

At each of the selected locations, both the CO(3-2) and CO(2-1) lines have quite similar if not identical spectral profiles, suggesting that they originate from (much) the same gas.  At locations corresponding to the center ($0\arcsec$,\,$0\arcsec$) and just east of the center ($1\farcs7$,\,$0\arcsec$), a blue wing is visually apparent in both CO(3-2) and CO(2-1).  This blue wing is more prominent relative to the main portion of the spectral profile, and furthermore extends to a much higher blueshifted velocity, in CO(3-2) than in CO(2-1).  At the approximate midpoint of the Inner filament ($-3\farcs5$,\,$0\arcsec$), the spectral profile in CO(2-1) exhibits a red wing that would be difficult to detect in CO(3-2) (there is a hint of its presence) if the ratio in brightness temperature between CO(3-2) and CO(2-1) is constant over the spectral profile.  At the eastern end ($-9\farcs5$,\,$1\farcs4$) and midpoint ($-13\farcs0$,\,$1\farcs4$) of the Western filament, the spectral profiles in both CO(3-2) and CO(2-1) appear similar.

To derive line ratios, we fitted Gaussians to the spectral profiles in CO(3-2) and CO(2-1) at each of the five selected locations.  The fitted Gaussians are shown in Figure\,\ref{line_profiles} for both CO(2-1) and CO(3-2), and their individual parameters listed in Table\,\ref{line parameters}.  We also list in this table, when both lines trace the same gas, the ratio in brightness temperature of CO(3-2) to CO(2-1), hereafter CO(3-2)/CO(2-1), computed from the brightness temperature ($T_{\rm B}$) integrated over the Gaussian(s) fitted to each line at each location.  At the center ($0\arcsec$,\,$0\arcsec$) and just east of the center ($1\farcs7$,\,$0\arcsec$), two Gaussian components are needed to provide a satisfactory fit to both the CO(3-2) and CO(2-1) spectral profiles; to obtain a reliable fit to the weaker component in CO(2-1) at the central position, we had to fix the central velocity of the brighter component to that of the brighter component found from fitting the CO(3-2) line.  At these two positions, the brighter component has a comparable width (at FWHM) in both CO(3-2) and CO(2-1), suggesting that this component traces the same gas in both lines.  This component has a line ratio of $0.87 \pm 0.14$ at the center, and $0.49 \pm 0.46$ just east of the center.  On the other hand, at each position, the dimmer component has a different central velocity and width in CO(3-2) compared with CO(2-1).  This component is not as well constrained in CO(2-1) as in CO(3-2), and furthermore does not extend to as high a blueshifted velocity in CO(2-1) as in CO(3-2) as mentioned earlier.  Although this blue wing therefore appears to have a ratio in brightness temperature between CO(3-2) and CO(2-1) well in excess of unity, caution must be applied when interpreting the line ratio of this component as the emission in CO(2-1) and CO(3-2) may not trace the same gas. 

At the approximate midpoint of the Inner filament ($-3\farcs5$,\,$0\arcsec$), two Gaussian components are needed to provide a satisfactory fit to the CO(2-1) but not CO(3-2) line.  As mentioned above, the CO(2-1) line exhibits a red wing that would not be readily detectable in CO(3-2) if the line ratio is constant over the spectral profile.  The brighter Gaussian component fitted to the main portion of the spectral profile in CO(2-1) has the same central velocity and width as the single Guassian fitted to the spectral profile in CO(3-2).  For this component, the line ratio is $0.68 \pm 0.19$.   At the remaining two locations both along the Western filament ($-9\farcs5$,\,$1\farcs4$ and $-13\farcs0$,\,$1\farcs4$), a single Gaussian provides a satisfactory fit to the spectral profiles in both CO(2-1) and CO(3-2).  At each of these locations, the fitted Gaussians have similar central velocities and widths in CO(2-1) and CO(3-2).  The line ratios at both these locations, $0.40 \pm 0.08$ and $0.44 \pm 0.12$, are comparable.  Evidently, the line ratio is higher at the center (close to 1.0) than further out (around 0.5).

Previous reports of CO line ratios in \object{NGC 1275} have been made with single-dish telescopes, where the individual filaments described above cannot be properly separated.  From observations in both CO(3-2) and CO(2-1) using the James Clerk Maxwell Telescope (JCMT), \citet{bri98} found $\rm CO(3-2)/CO(2-1) \approx 1.0$ in both 14\arcsec\ (5.0\,Kpc) and 21\arcsec\ (7.6\,Kpc) beams (as measured at FWHM) towards the center of \object{NGC 1275}.  By comparing their observations with those in CO(1-0) by \citet{reu93} using the IRAM 30-m telescope, they found $\rm CO(2-1)/CO(1-0) \approx 0.7$ in a 21\arcsec\ (7.6\,Kpc) beam towards the center of \object{NGC 1275}.    From observations of different regions (extending far beyond the field of view of that mapped here in CO(3-2)) in both CO(1-0) and CO(2-1) with the IRAM 30-m telescope, 
\citet{sal08b} measured CO(2-1)/CO(1-0) line ratios (convolved to a common angular resolution of 22\arcsec, or a spatial resolution 7.9\,kpc) that exceed unity in most of the regions observed, and which overall span the wide range $\sim$0.5--2.4.  In contradiction with \citet{bri98}, \citet{sal08b} found $\rm CO(2-1)/CO(1-0) \approx 1.7$ (i.e., much higher than unity) towards the center of \object{NGC 1275}.  As emphasized by \citet{sal08b}, the spectra in CO(1-0) towards the center of \object{NGC 1275} is badly affected by baseline ripples caused by standing waves between the dish and subreflector set up by the strong AGN continuum emission of \object{NGC 1275} at 3~mm.  The AGN continuum emission becomes progressively weaker towards shorter wavelengths, and does not appreciably affect the spectra in higher CO transitions.  As a consequence, the CO(2-1)/CO(1-0) line ratio measured by both \citet{bri98} and \citet{sal08b} towards the center of \object{NGC 1275} may not be reliable.  More than one CO(1-0) beam away from the center of the galaxy, \citet{sal08b} measure CO(2-1)/CO(1-0) line ratios that can have values close to as well as significantly smaller or larger than unity (i.e., spanning the range $\sim$0.5--2.4 as mentioned above).

\section{Excitation of Trace Molecules}\label{CO Excitation}
Interstellar molecules can be excited by collisions with either neutral (e.g., H, He, and H$_2$) or charged (e.g., electron) particles, as well as by absorption of radiation (e.g., the Cosmic Microwave Background, emission from dust mixed in with the molecular gas, as well as emission from the molecules themselves).  As mentioned in $\S\ref{Introduction}$, \citet{fab11} suggest that the surrounding X-ray emitting gas that penetrates into the filamentary nebula of \object{NGC 1275} constitutes the source of energetic particles advocated by \citet{fer09} for exciting, through the production of suprathermal secondary electrons, the atomic and ionic gas components that emit in optical and infrared lines, as well as the molecular gas component that emits in H$_2$ ro-vibrational infrared lines.  In the following, we first consider in detail whether collisions with electrons is a viable mechanism for exciting the trace molecules (i.e., CO, HCN, HCO$^+$, and CN) detected from \object{NGC 1275}, before turning to more conventional sources of excitation.

\subsection{Electron-impact excitation}\label{Electron-impact excitation}

Assuming long-range interactions only, \citet{dic75} computed the theoretical cross-sections for $|\Delta{J}| = 1$ rotational transitions of linear polar molecules due to electron impact.  In more modern calculations incorporating also short-range interactions, the dominance of $|\Delta{J}| = 1$ over $|\Delta{J}| > 1$ transitions have been confirmed for all (strongly) polar linear molecules so far considered \citep{fau07a,fau07b,var10}.  Molecular ions with dipole moments smaller than $\sim$2\,D, however, can have stronger $|\Delta{J}| > 1$ than $|\Delta{J}| = 1$ transitions \citep{fau01}.  To compute the corresponding molecular excitation rate coefficients due to electron impact, the velocity distribution of the impacting electrons must be known.  A thermal (Maxwell-Boltzmann) velocity distribution for the electrons is normally assumed (as is the case in all the references cited above); here, we make the same assumption for the electrons produced through ionization of the ISM in \object{NGC 1275}, as would be the case if these electrons thermalize rapidly due to collisions.  From Eq.\,(2.21) of \citet{dic75}, we find electron-impact excitation rate coefficients for CO\,$J = 0 \rightarrow 1$, $1 \rightarrow 2$, and $2 \rightarrow 3$ as plotted in Figure\,\ref{CO excitation rate}.  Over the temperature range 10--1000\,K for the electrons (note that the electron temperature can, in principle, be higher than the gas temperature), these rate coefficients change by only a factor of $\sim$2 or less.
The rate coefficients for electron-impact excitation of CO at the transitions shown in Figure\,\ref{CO excitation rate} are a factor of $\sim$$10^{2}$
larger than the corresponding values for H$_2$-impact excitation \citep[for the most recent theoretical calculations of H$_2$-impact excitation rate coefficients for CO, see][]{yan10}, explaining why we need to properly evaluate the likelihood of electron-impact excitation of CO in an environment with copious electrons.

For electrons to compete with or dominate over H$_2$ in exciting the CO(1-0), CO(2-1), and CO(3-2) lines, the ionization fraction of the (predominantly) H$_2$ gas must therefore be of order $10^{-2}$ or higher.  (Note that the ionization fraction, $x(\rm e)$, of molecular gas is defined here as $x_{\rm e} = n({\rm e}) / 2 \, n(\rm H_2)$, where $n(\rm H_2)$ the number density of $\rm H_2$ molecules.  This definition is the same as that used by \citet{lis12}, thus providing a direct comparison with his computations for molecular line emission from diffuse partially-molecular clouds in our Galaxy.)
Such ionization fractions are many orders of magnitude higher than those of giant molecular clouds (GMCs) in our Galaxy, where typical estimates range from $\sim$$10^{-9}$--$10^{-6}$ \citep[e.g.,][and references therein]{cas02,flo07}, and about two
orders of magnitude higher than those of diffuse partially-molecular clouds in our Galaxy, where the ionization fraction is estimated to be $\sim$1--$2 \times 10^{-4}$ \citep[e.g.,][and references therein]{lis12}.  In \object{NGC 1275}, assuming that the different gas phases of the ISM are in pressure equilibrium with the surrounding ICM X-ray gas, which has $n_e T \approx 10^{6.5} {\rm \ cm^{-3} \ K}$ where $n_e$ is the electron density and $T$ the gas temperature, \citet{fer09} find that the mid-IR H$_2$ lines in a particularly well-studied (bundle of) filament(s) originates from predominantly molecular
regions where the total hydrogen ($\rm H_{total} = H^{+}+H+H_2$) gas density is $\sim$$10^{5.2} {\rm \ cm^{-3}}$ and temperature $\sim$$10^{1.3}$\,K ($\sim$20\,K).  In the model of \citet{fer09}, such regions have ionization fractions (far) below $10^{-5}$ so that CO excitation by electrons is negligible compared with H$_2$.  Because the H$_2$ density in these region exceeds by 1--2 orders of magnitude the critical densities for H$_2$-impact excitation of CO(1-0), CO(2-1), and CO(3-2), collisional excitation by H$_2$ alone is sufficient for CO to be thermalized in all these transitions.  In the model of \citet{fer09}, 
an ionization fraction as high as 
$\sim$$10^{-2}$ is reached only in regions where the total hydrogen gas density is $\sim$$10^{3.5} {\rm \ cm^{-3}}$ and temperature $\sim$$10^{3}$\,K; in such regions, however, the gas is no longer primarily molecular (H$_2$ abundance of only $\sim$0.01 or less).  Note that both the electron-impact cross-section and dissociation threshold for CO \citep{cos93} are similar with those for H$_2$ \citep[e.g.,][]{dal99}, and so CO should just be as strongly dissociated as H$_2$; i.e., we need not be concerned with the possibility that CO survives electron-impact dissociation while H$_2$ does not, giving rise to regions where CO is excited by electrons in primarily atomic gas.

A deeper appreciation of the importance electrons might play in exciting CO comes from comparing the electron density with the critical density.\footnote{Particle density whereby the collisional de-excitation rate is equal to the spontaneous emission rate (the Einstein A coefficient), providing a rough measure \citep[although caution is required; e.g., see][]{shi15} of whether a particular molecule is strongly excited at a given transition.} for electron-impact excitation, as shown in Figure\,\ref{CO excitation rate}.  The latter is $\sim$10--$30 {\rm \ cm^{-3}}$ for CO(1-0), $\sim$90--$270 {\rm \ cm^{-3}}$ for CO(2-1), and $\sim$350--$980 {\rm \ cm^{-3}}$ for CO(3-2) over the electron temperature range 10--1000\,K.  A comparison with the electron density in the ICM X-ray gas surrounding the line-emitting filaments in \object{NGC 1275} is instructive.
The electron density of the ICM at the core of the Perseus cluster is $\sim$$0.04 {\rm \ cm^{-3}}$, where the electron energy is $\sim$3\,KeV \citep{chu02}.  Assuming that this X-ray gas penetrates at full efficiency (i.e., efficiency of $\sim$100\%) into the filaments, and that each electron of this X-ray gas (with an energy of $\sim$3\,KeV) produces (at most) $\sim$80 secondary electrons \citep{dal99}, the secondary electron density in the filaments can be as high as $\sim$$3 {\rm \ cm^{-3}}$.  Such electron densities are about 2 orders of magnitude below the critical densities for electron-impact excitation of CO(2-1) and CO(3-2), and up to an order of magnitude below the critical density for electron-impact excitation of CO(1-0).  If instead the efficiency at which the X-ray gas penetrates into the filaments is only $\sim$10\% as proposed by \citet{fab11}, then the secondary electron density becomes at most $\sim$$0.3 {\rm \ cm^{-3}}$, which is at least an order of magnitude below the critical density for electron-impact excitation of CO(1-0).

In addition to ionization by the X-ray gas that penetrates into the filaments, photoionization by the surrounding X-ray gas also contributes to the ionization state in the filaments.  Thus, the electron density in the filaments may be significantly higher than that produced solely through ionization by the infiltrating X-ray gas as computed above.  At this stage, we note that, as pointed out by \citet{lim12}, the X-ray gas has a surface brightness just below that of the filaments (dominated in brightness by its optical emission lines) within a central radius of $\sim$20\,kpc, and drops increasingly below the surface brightness of these filaments towards larger radii.  Thus, photoionization by the surrounding X-ray gas is not by itself sufficiently powerful to excite the optical emission-lines from \object{NGC 1275}, with the deficiency becoming especially apparent beyond a central radius of $\sim$20 kpc.  


In summary, in predominantly molecular gas, an ionization fraction of $\sim$$10^{-2}$ is required for electrons to compete with H$_2$ for exciting CO so as to produce CO(1-0), CO(2-1), and CO(3-2).  At such a high ionization fraction, however, the gas is no longer predominantly molecular \citep[H$_2$ abundance $\lesssim 0.01$ in the model of][]{fer09}. 
Based on the arguments presented above, we therefore conclude that electron-impact excitation cannot dominate over H$_2$ in exciting CO in \object{NGC 1275}.
The same conclusion has been reached for CO in Galactic gas clouds, including diffuse partially-molecular clouds \citep[see][]{lis12} where the ionization fraction is several orders of magnitude higher than that of GMCs.

\citet{sal08b} have reported the detection of HCN(3-2), and \citet{bay11} the detection of HCO$^+$(3-2), CN(2-1), and tentatively also C$_2$H(3-2), at millimeter wavelengths towards \object{NGC 1275}.  The critical densities for H$_2$-impact excitation of all these lines are 2--3 orders of magnitude higher than the corresponding values for CO.  In Galactic molecular clouds, these lines originate from dense condensations (referred to as cores) --- in which stars form --- embedded within more diffuse molecular gas traced by CO, the latter constituting the bulk in mass and volume of a cloud.  Yet, apart from two compact HII-regions so far identified \citep{shi90,lim12}, neither of which are coincident with detectable molecular gas as traced in CO \citep{lim08,hol09,lim12}, there is no evidence (e.g., optical spectra characteristic of HII regions, or blue light from young star clusters) for recent star formation within the line-emitting filaments in \object{NGC 1275}.  (Instead, as shown in Figure\,\ref{NGC 1275}, recent star formation can be found away from, albeit sometimes closely associated with, the line-emitting filaments.)  We therefore also consider whether suprathermal electrons play a role in exciting HCN, HCO$^+$, and CN in relatively diffuse molecular gas in \object{NGC 1275}.

For collisions with electrons, the cross-section of $|\Delta{J}| = 1$ rotational transitions of linear polar molecules is proportional to the square of the molecule's dipole moment \citep{dic75}.  Thus, the importance of electron-impact excitation in molecular clouds can be far greater for molecules such as CN (dipole moment of 1.45\,D), HCN (2.98\,D), and HCO$^+$ (3.90\,D) compared with CO (0.11\,D), an issue recognized as far back as \citet{dic77}.  This point has been emphasized by \citet{lis12} for diffuse partially-molecular clouds in our Galaxy, from which HCN(1-0) and HCO$^+$(1-0) have been detected despite such clouds having a density of $< 10^{3} {\rm \ cm^{-3}}$.  As an instructive illustration, let us consider HCN.  The critical density for H$_2$-impact excitation of the HCN(1-0) line is $\sim$$10^6 {\rm \ cm^{-3}}$, and explains why HCN is normally considered to trace high-density molecular gas characteristic of dense cores in which stars form.  
From Eq.(2.21) of \citet{dic75}, we find electron-impact excitation rate coefficients for HCN\,$J = 0 \rightarrow 1$, $1 \rightarrow 2$, and $2 \rightarrow 3$ as plotted in Figure\,\ref{HCN excitation rate} (solid lines).  We also plot in Figure\,\ref{HCN excitation rate} (dotted lines) the same electron-impact excitation rate coefficients from the more recent computations by \citet{fau07a}.
Over the temperature range 10--1000\,K for the electrons, these rate coefficients change by only a factor of $\sim$2 or less.  The rate coefficients for electron-impact excitation of HCN at the transitions shown in Figure\,\ref{HCN excitation rate} are a factor of $\sim$$10^{5}$--$10^{6}$ larger than the corresponding rate coefficients for H$_2$-impact excitation.  Thus, in gas clouds where the ionization fraction is equal to or exceeds $\sim$$10^{-6}$--$10^{-5}$, electrons compete with or dominate over H$_2$ as the primary source of HCN excitation (by comparison, as shown above, an ionization fraction of at least $10^{-2}$ is required for electrons to compete with H$_2$ for exciting CO).  Indeed, in Galactic diffuse partially-molecular clouds where the bulk H$_2$ density is just $\sim$15--$250 {\rm \ cm^3}$ but the ionization fraction $\sim$1--$2 \times 10^{-4}$, excitation by electrons rather than H$_2$ produce the HCN(1-0) and HCO$^+$(1-0) lines observed in emission from these clouds \citep[e.g.,][]{lis12}.  In an environment with copious electrons, line emission from molecules with high dipole moments such as CN, HCN, HCO$^+$ should therefore not be automatically assumed to trace very dense molecular gas characteristic of star-forming sites in Galactic GMCs.

In the model of \citet{fer09}, ionization fractions of $10^{-6}$--$10^{-5}$, and indeed up to $\sim$$10^{-4}$ if not higher (as high as nearly $10^{-3}$), are easily reached where the gas is primarily molecular (H$_2$ abundance $>0.5$).  
It is therefore entirely possible that the HCN(3-2), HCO$^+$(3-2), and CN(2-1) lines detected from \object{NGC 1275} are excited by collisions with electrons rather than H$_2$; i.e., in regions where the H$_2$-gas density is far below the critical densities for H$_2$-impact excitation of these transitions, corresponding to $\sim$$10^{6} {\rm \ cm^{-3}}$ for HCO$^+$(3-2) and $\sim$$10^{7} {\rm \ cm^{-3}}$ for HCN(3-2) and CN(2-1).   
This possibility should give pause before attributing rotational transitions from these molecules to emission from regions with very dense ($>10^{5} {\rm \ cm^{-3}}$) H$_2$ gas, as has so far been assumed to be the case in \object{NGC 1275} \citep{sal08b,bay11}.  Instead, these lines may be excited by electrons in molecular gas with relatively high ionization fractions of up to $\sim$$10^{-4}$ (comparable with that of diffuse partially-molecular clouds in our Galaxy) if not higher, rather than in dense molecular gas.

Figure\,\ref{HCN excitation rate} shows the critical densities for electron-impact excitation of HCN as calculated from the collisional de-excitation rates given by \citet{dic75} (solid lines) and \citet{fau07a}(dotted lines).  Based on the more recent computations by \citet{fau07a}, we find critical densities of $\sim$6--$14 {\rm \ cm^{-3}}$ for HCN(1-0), $\sim$60--$130 {\rm \ cm^{-3}}$ for HCN(2-1), and $\sim$250--$470 {\rm \ cm^{-3}}$ for HCN(3-2) over the electron temperature range 10--1000\,K .  These critical densities are comparable to, at best only a factor of $\sim$2 smaller than, the corresponding critical densities for electron-impact excitation of the CO(1-0) to CO(3-2) lines.  (Although the electron-impact excitation rate coefficients for HCN are 3 orders of magnitude higher than that for CO, the Einstein A coefficent also is nearly 3 orders of magnitude higher for HCN than for CO between the same set of rotational levels.)  In $\S\ref{Gas HCN}$, we explore quantitatively whether electron-impact excitation can produce the observed HCN(3-2) emission from NGC\,1275 in relatively diffuse molecular gas (i.e., much lower in density than the critical density for H$_2$-impact excitation of $\gtrsim 10^6 {\rm \ cm^{-3}}$).

Finally, we remark on why HCO$^+$(3-2) is so much (over ten times) brighter, and CN(2-1) also much (over four times) brighter, than HCN(3-2) as observed towards the center of \object{NGC 1275}.  \citet{bay10} find that chemical reactions in molecular gas subjected to relatively high ionization rates can dramatically alter the fractional abundances (measured with respect to H$_2$) of some molecules: the fractional abundances of both HCO$^+$ and CN increase dramatically, whereas the fractional abundance of HCN decreases dramatically, with the ionization rate.  Such changes in abundances might explain why both HCO$^+$(3-2) and CN(2-1) are much brighter than HCN(3-2) as observed towards the center of \object{NGC 1275}.  Here we point out an additional reason, which is especially relevant if these molecules are excited by electrons rather than H$_2$.  
The latest computations of electron-impact excitation rate coefficients for HCO$^+$ by \citet{fau07b} indicate not just higher $|\Delta{J}| = 1$ excitation rate coefficients in keeping with the larger dipole moment of this molecule compared with HCN, but also much higher $|\Delta{J}| > 1$ (e.g., $J = 0 \rightarrow 2$ and $J = 0 \rightarrow 3$) excitation rate coefficients compared with HCN.  The critical density for electron-impact excitation of HCO$^+$ for transitions higher than the ground-state transition is therefore significantly lower than the corresponding critical density for electron-impact excitation of HCN (as well as CN), and may therefore also contribute towards explaining the large difference in brightness between these two lines in their $J = 3 \rightarrow 2$ transitions.  The same issue has been pointed out by \citet{lis12} for diffuse partially-molecular clouds, where HCO$^+$(1-0) is predicted (and observed) to be stronger than HCN(1-0). 

\subsection{Radiative excitation}\label{Radiative excitation}
Molecules can be excited by radiation at millimeter to infrared wavelengths from sources such as the Cosmic Microwave Background (CMB).   Apart from the CMB, molecules also can be excited by radiation from dust (which emits primarily in the infrared) mixed in with the molecular gas, as well as radiation from the molecules themselves (i.e., self absorption of the radiation, referred to as radiative trapping).  Although radiative trapping, in particular, can contribute appreciably to exciting molecules at high optical depths \citep[e.g., see][]{shi15}, in Galactic GMCs collisions with H$_2$ dominate the excitation of CO.  Using {\it Herschel}, \citet{mit12} detected far-IR emission from dust in the filaments at the inner regions of \object{NGC 1275}.  They infer two bulk (albeit highly uncertain) temperatures for this dust, one at about 30--40\,K and the other at about 100\,K.  The cooler component has an inferred mass somewhat less than $10^7 {\rm \ M_{\sun}}$, and the warmer component somewhat less than $10^5 {\rm \ M_{\sun}}$.  Assuming a Galactic conversion between CO luminosity and H$_2$-gas mass, the total molecular-gas-to-dust ratio is of order 5000-10000, well over one and reaching two orders of magnitude higher than that in Galactic GMCs.  Thus, unless the mass in molecular gas is overestimated by many orders of magnitude, radiation from dust cannot play a significant role in exciting the molecular gas.

As described in $\S$\ref{line ratio}, the line ratio CO(3-2)/CO(2-1) reaches its highest value at and close to the center of \object{NGC 1275}.  Here, intense millimeter to infrared radiation from the AGN, as well as radiation from dust heated by optical/UV radiation from the AGN, can potentially contribute to the excitation of CO as well as other trace molecules.

\subsection{H$_2$ collisional excitation}\label{H$_2$ collisional excitation}
As mentioned above, collisions with H$_2$ are believed (and universally assumed) to dominate the excitation of trace molecules in both Galactic GMCs and extragalactic molecular clouds.  From the aforementioned discussion, we find that CO in \object{NGC 1275} also is excited by collisions with H$_2$.  On the other hand, for molecules with significantly larger dipole moments such as CN, HCN, and HCO$^+$, electrons rather than H$_2$ can be the primary source of excitation, as is the case in diffuse partially-molecular clouds in our Galaxy.

\section{Physical Conditions of Molecular Gas}\label{physical conditions}

\subsection{Gas emitting in CO}\label{Gas CO}

As explained in $\S\ref{Electron-impact excitation}$, a relatively high ionization fraction of $\sim$$10^{-2}$ is required for electrons to compete with H$_2$ in collisionally exciting CO, so high that the gas is no longer primarily molecular (H$_2$ abundance of $\lesssim 0.01$) but instead atomic.  
Close to the center, continuum radiation from the AGN can potentially also contribute, either directly or through reprocessed radiation from heated dust, to exciting CO.  Without knowledge of any such contribution, in deriving below the physical properties of the molecular (primarily H$_2$) gas traced in CO, we consider only collisional excitation with H$_2$, radiative excitation from the cosmic microwave background (CMB), and self-absorption of radiation produced by the gas itself (radiative trapping).

For CO\,$J=2$ to be thermalized (i.e., to have an excitation temperature, $T_{\rm ex}$, equal to the gas kinetic temperature, $T_{\rm k}$), the H$_2$ gas density must be close to or exceed a critical density \citep[see][]{yan10} of $n_{\rm crit} \sim 5 \times 10^3 {\rm \ cm^{-3}}$ at $T_{\rm k} \sim 10$\,K, decreasing gradually to $n_{\rm crit} \sim 2 \times 10^3 {\rm \ cm^{-3}}$ at $T_{\rm k} \sim 100$\,K (and beyond).  The critical density for the CO\,$J=3$ level is a factor of a few larger, $n_{\rm crit} \sim 1 \times 10^4 {\rm \ cm^{-3}}$ at $T_{\rm k} \sim 10$\,K decreasing gradually to $n_{\rm crit} \sim 8 \times 10^3 {\rm \ cm^{-3}}$ at $T_{\rm k} \sim 100$\,K.  In Figure\,\ref{line_ratio_theory}, we plot the predicted ratio in brightness temperature between CO(3-2) and CO(2-1) --- CO(3-2)/CO(2-1) --- as a function of H$_2$ gas density, $n({\rm H_2})$, and CO column density per unit velocity, $N_{\rm {CO}}$/$\Delta v$, at six different gas temperatures of $T_{\rm k} = 10$\,K, 20\,K, 30\,K, 50\,K, 100\,K, and 200\,K.  Notice that the H$_2$ gas density required to attain a given line ratio decreases as the gas temperature increases, reflecting the gradual decrease in $n_{\rm crit}$ as $T_{\rm k}$ increases.  These computations were made using own our code, which uses the large velocity gradient approximation (LVG) for the escape probability of photons from a given parcel of gas.  Molecular data for the CO molecule, including its energy levels, radiative transition rates, and collisional cross sections, were taken from the Leiden Atomic and Molecular Database \citep{sch05}. We include in our calculations all rotational levels up to $J=15$.  The temperature of the CMB radiation field is set to 2.73 K.  The statistical equilibrium equations containing the contribution of the radiation field calculated within the LVG approximation are solved iteratively using the Newton-Raphson method.  Convergence is achieved when the relative change in level populations is less than 10$^{-3}$ between successive iterations. The inputs of our code are the gas kinetic temperature ($T_{\rm k}$), the molecular hydrogen density ($n({\rm H_2})$), and the CO column density per unit velocity ($N_{\rm {CO}}$/$\Delta v$).

At relatively low values of $N_{\rm {CO}}$/$\Delta v$ where the curves representing different line ratios are closely horizontal, the CO(2-1) and CO(3-2) lines are optically thin (i.e., optical depth $\tau \lesssim 1$).  In this regime, both lines are subthermally excited; at a given $N_{\rm {CO}}$/$\Delta v$, as $n({\rm H_2})$ increases the CO\,$J=3$ level becomes increasingly well populated, thus resulting in a higher $\rm CO(3-2)/CO(2-1)$.  Values $\rm CO(3-2)/CO(2-1) \geq 1$ (above and to the left of the dotted line, which corresponds to $\rm CO(3-2)/CO(2-1) = 1$) require both lines to be optically thin, and so are restricted to relatively low $N_{\rm {CO}}$/$\Delta v$ values.  As $N_{\rm {CO}}$/$\Delta v$ increases, both lines become more optically thick and radiative excitation through self-absorption (i.e., radiative trapping) contributes increasingly to the rise in their excitation temperatures.  As a consequence, to attain a given line ratio in the regime $\rm CO(3-2)/CO(2-1) < 1$ (i.e., below and to the right of the dotted line), $n({\rm H_2})$ decreases as $N_{\rm {CO}}$/$\Delta v$ increases.  Equivalently, at a constant $n({\rm H_2})$ in this regime, the line ratio increases as $N_{\rm {CO}}$/$\Delta v$ increases.  When $n({\rm H_2})$ and $N_{\rm {CO}}$/$\Delta v$ reach values where both CO(2-1) and CO(3-2) are thermalized and optically thick ($\tau \gg 1$), the line ratio reaches a value of just below unity (region to the right of the dotted line encompassed by $\rm 0.9 < CO(3-2)/CO(2-1) < 1.0$).

Figure\,\ref{line_ratio_theory} shows that, to attain a line ratio CO(3-2)/CO(2-1) as high as that measured of $\sim$0.9, the gas temperature $T_{\rm k} \gtrsim 20 {\rm \ K}$.  Given the environment in which the molecular gas in cluster central elliptical galaxies is immersed (i.e., irradiation by the surrounding X-ray gas, infiltration of the X-ray gas into the molecular gas), it should come as no surprise if the temperature of the bulk molecular gas in these galaxies is higher than that of Galactic GMCs.  To further constrain the relevant range in physical parameters for the molecular gas, we note that the model of \citet{fer09} assumes the different gas components of the ISM are in thermal pressure equilibrium with each other as well as the surrounding X-ray gas.  The pressure of the X-ray gas towards the inner regions of \object{NGC 1275} is $n_e T \sim 10^{6.5} {\rm \ cm^{-3} \ K}$ \citep{san07}.  For temperatures over the range $\sim$20--100\,K for the molecular gas, pressure equilibrium through thermal support alone requires $n(\rm H_2) \sim 10^{4.5} {\rm \ cm^{-3}}$ (for $T_{\rm k} = 100 {\rm \ K}$) to $10^{5.2} {\rm \ cm^{-3}}$ (for $T_{\rm k} = 20 {\rm \ K}$).  Such high gas densities are incompatible with the range in line ratios measured of $\sim$0.4--0.9; to obtain a line ratio as low as 0.4 at $T_{\rm k} \gtrsim 20 {\rm \ K}$, $n(\rm H_2) \lesssim 10^4 {\rm \ cm^{-3}}$.  \citet{hol09} argue that to be supported against collapse and therefore not exhibit vigorous star formation, the molecular gas in the filaments must be supported predominantly by (internal) turbulence or threaded by magnetic fields (see more detailed discussion in $\S\ref{stability filaments}$).  Similarly, \citet{fab08} argue that strong magnetic fields are required to stabilize thread-like structures observed in the optical line-emitting filaments at high angular resolutions.  If so, depending on the amount of support provided by turbulence and/or magnetic fields, the density of the molecular gas can be considerably lower than that computed above assuming only thermal support.  To constrain $N_{\rm {CO}}$/$\Delta v$, we note that the CO-emitting filaments are not resolved in width, and therefore have a cross-sectional diameter of less than $\sim$1\,kpc.  The latter places an upper limit on the CO column density, assuming a fractional abundance $\rm [CO/H_2] \sim 10^{-5}$ as in Galactic molecular clouds, of $N_{\rm {CO}} < 10^{16} \ n(\rm H_2) {\rm \ cm}$.  For $n(\rm H_2) \lesssim 10^4 {\rm \ cm^{-3}}$ as is required to reproduce the observed line ratios, and measured CO linewidths (at FWHM) over a synthesized beam spanning the range $\sim$50--$100 {\rm \ km \ s^{-1}}$ (Table\,\ref{line parameters}), $N_{\rm {CO}}/\Delta v \lesssim 10^{18} {\rm \ cm^{-2} \ km^{-1} \ s}$.  

In Figure\,\ref{CO model 1}, we plot the optical depths, excitation temperatures (reflecting the level distributions), and brightness temperatures of the CO(1-0), CO(2-1), and CO(3-2) lines as a function of H$_2$ gas density for $T_{\rm k} = 30$\,K (left column), $T_{\rm k} = 50$\,K (middle column), and $T_{\rm k} = 100$\,K (right column).  We assume in all these plots that $N_{\rm {CO}}/\Delta v = 3 \times 10^{16} {\rm \ cm^{-2} \ km^{-1} \ s}$, which is at or close to the knee of the curve for $\rm CO(3-2)/CO(2-1) \sim 0.9$  in Figure\,\ref{line_ratio_theory} over the selected range of temperatures; i.e., at or close to the maximum possible value of $N_{\rm {CO}}/\Delta v$ for which $n(\rm H_2)$ is at or close to its highest value.  As is apparent in Figure\,\ref{CO model 1}, the optical depth in CO(3-2) initially increases as $n(\rm H_2)$ rises because the $J=3$ level becomes increasingly well populated (at the expense of lower $J$ levels).  Eventually, the $J \geq 4$ level becomes increasingly well populated at the expense of the $J \leq 3$ level, resulting in a decrease in the optical depth of CO(3-2).  By comparison, over the entire range plotted for $n(\rm H_2)$, the optical depths in both CO(1-0) and CO(2-1) fall with increasing $n(\rm H_2)$ as both the $J=1$ and $J=2$ levels become less well populated in preference to the higher $J$ levels.  As is apparent from Figure\,\ref{line_ratio_theory}, to attain $\rm CO(3-2)/CO(2-1) \sim 0.9$ at $N_{\rm {CO}}/\Delta v = 3 \times 10^{16} {\rm \ cm^{-2} \ km^{-1} \ s}$, $n(\rm H_2) \sim 10^{4.2} {\rm \ cm^{-3}}$ at $T_{\rm k} = 30$\,K, decreasing to $n(\rm H_2) \sim 10^{3.8} {\rm \ cm^{-3}}$ at $T_{\rm k} = 50$\,K, and further decreasing to $n(\rm H_2) \sim 10^{3.5} {\rm \ cm^{-3}}$ at $T_{\rm k} = 100$\,K.  Such densities imply a corresponding depth along the line-of-sight for the CO-emitting gas of about $\sim$10--30\,pc, towards the lower bound in size of Galactic GMCs.  Assuming cylindrical filaments observed from the side (i.e., diameters in the range 10--30\,pc), the filling factor in a synthesized beam is therefore about 0.01--0.03 ($\sim$$r_{\rm c}/r_{\rm s}$, where $r_{\rm c}$ is the radius of the cylindrical filament and $r_{\rm s} \approx 0.5 {\rm \ kpc}$ the radius of the synthesized beam).  From Figure\,\ref{CO model 1}, the predicted peak brightness temperatures in both CO(2-1) and CO(3-2) is $T_{\rm B} \sim 20$\,K for $n(\rm H_2) \sim 10^{4.2} {\rm \ cm^{-3}}$ at $T_{\rm k} = 30$\,K, $T_{\rm B} \sim 25$\,K for $n(\rm H_2) \sim 10^{3.8} {\rm \ cm^{-3}}$ at $T_{\rm k} = 50$\,K, and $T_{\rm B} \sim 25$\,K for $n(\rm H_2) \sim 10^{3.5} {\rm \ cm^{-3}}$ at $T_{\rm k} = 100$\,K.  By comparison, the measured peak brightness temperatures of both lines towards the center of \object{NGC 1275}, where $\rm CO(3-2)/CO(2-1) \sim 0.9$, is about 0.4\,K.  Taking into account the implied filling factor of about 0.01--0.03 and hence an actual brightness temperature of about 10--40\,K, our model predictions are therefore in good accord with the measurements.  

For values lower than $N_{\rm {CO}}/\Delta v = 3 \times 10^{16} {\rm \ cm^{-2} \ km^{-1} \ s}$, $n(\rm H_2)$ remains approximately constant for $\rm CO(3-2)/CO(2-1) \sim 0.9$ over the range of temperatures considered.  As a consequence, as $N_{\rm {CO}}/\Delta v$ decreases below the aforementioned value, the depth of the CO-emitting gas and hence also its filling factor (assuming a cylindrical filament seen from the side) must decrease in nearly direct proportion.  A drop in $N_{\rm {CO}}/\Delta v$ therefore results in a drop in the optical depth and hence excitation temperatures (less radiative trapping) of both the CO(2-1) and CO(3-2) lines. Combined also with the drop in filling factor, the predicted brightness temperatures of both these lines increasingly depart from (becomes much lower than) those measured with decreasing $N_{\rm {CO}}/\Delta v$.  For values higher than $N_{\rm {CO}}/\Delta v = 3 \times 10^{16} {\rm \ cm^{-2} \ km^{-1} \ s}$, $n(\rm H_2)$ decreases progressively with increasing $N_{\rm {CO}}/\Delta v$ for $\rm CO(3-2)/CO(2-1) \sim 0.9$.  As a consequence, the depth of the CO-emitting gas and hence also its filling factor must increase with $N_{\rm {CO}}/\Delta v$.  The increase in filling factor results in a predicted brightness temperatures for both CO(2-1) and CO(3-2) that increasingly depart from (becomes much higher than) those measured with increasing $N_{\rm {CO}}/\Delta v$.  The tension between increasing and decreasing $N_{\rm {CO}}/\Delta v$ from the knee of the curve for $\rm CO(3-2)/CO(2-1) \sim 0.9$ means that $N_{\rm {CO}}/\Delta v \sim 3 \times 10^{16} {\rm \ cm^{-2} \ km^{-1} \ s}$ and $n(\rm H_2) \sim 10^{4} {\rm \ cm^{-3}}$ are required to match both the observed line ratio and measured brightness temperatures of the brighter component in the CO(2-1) and CO(3-2) lines towards the nucleus.  As mentioned above, these requirements translate to a depth for the CO-emitting gas of about 10--30\,pc and a filling factor (assuming a cylindrical filament seen from the side) of about 0.01--0.03 for a gas temperature of 30--100\,K.

Towards the Western filament, $\rm CO(3-2)/CO(2-1) \sim 0.4$.  For this line ratio, at most $n(\rm H_2) \sim 10^3 {\rm \ cm^{-3}}$ over the range of gas temperatures from $\sim$30--100\,K as can be seen in Figure\,\ref{line_ratio_theory}.  Assuming as before a cylindrical filament with a diameter of $\sim$10\,pc seen from the side, at most $N_{\rm {CO}}/\Delta v \sim 3 \times 10^{15} {\rm \ cm^{-2} \ km^{-1} \ s}$, which is close to the knee of the curve for $\rm CO(3-2)/CO(2-1) \sim 0.4$.  Under these conditions, we find that we cannot reproduce the observed brightness temperature given an implied filling factor in a synthesized beam of $\sim$0.01 (the predicted brightness temperature being an order of magnitude too low).  Instead, we require factor of a few higher value of $N_{\rm {CO}}/\Delta v \sim 1 \times 10^{16} {\rm \ cm^{-2} \ km^{-1} \ s}$, which for $\rm CO(3-2)/CO(2-1) \sim 0.4$ requires $n(\rm H_2) \sim 10^{2.9} {\rm \ cm^{-3}}$ at $T_{\rm k} = 30$\,K, decreasing to $n(\rm H_2) \sim 10^{2.6} {\rm \ cm^{-3}}$ at $T_{\rm k} = 50$\,K, and further decreasing to $n(\rm H_2) \sim 10^{2.3} {\rm \ cm^{-3}}$ at $T_{\rm k} = 100$\,K.  The corresponding depth is $\sim$40\,pc for $n(\rm H_2) \sim 10^{2.9} {\rm \ cm^{-3}}$, $\sim$80\,pc for $n(\rm H_2) \sim 10^{2.9} {\rm \ cm^{-3}}$, and $\sim$170\,pc for $n(\rm H_2) \sim 10^{2.3} {\rm \ cm^{-3}}$, towards the upper bound in size of Galactic GMCs.  The corresponding filling factor ranges from $\sim$0.04 for $n(\rm H_2) \sim 10^{2.9} {\rm \ cm^{-3}}$ to $\sim$0.17 for $n(\rm H_2) \sim 10^{2.3} {\rm \ cm^{-3}}$.  In Figure\,\ref{CO model 2}, we plot the optical depths, excitation temperatures, and brightness temperatures of the CO(1-0), CO(2-1), and CO(3-2) lines as a function of H$_2$ gas density for $T_{\rm k} = 30$\,K (left column), $T_{\rm k} = 50$\,K (middle column), and $T_{\rm k} = 100$\,K (right column), but now for $N_{\rm {CO}}/\Delta v = 1 \times 10^{16} {\rm \ cm^{-2} \ km^{-1} \ s}$.  The predicted peak brightness temperatures in CO(2-1) is $T_{\rm B} \sim 7$--8\,K at the different aforementioned densities over the range of temperatures considered.  For comparison, the peak brightness temperature measured in CO(2-1) towards the Western filament is $\sim$0.3\,K.  Taking into account the implied filling factor in this case of about 0.04--0.17 and hence an actual brightness temperature of about 2--8\,K, our model predictions are therefore in good accord with the measurements.

Our computations as outlined above indicates filaments having smaller radii towards the center than those further out.  This behavior could be produced by tidal stretching of the filaments.  Alternatively, if the filaments have the same radii, then towards the center radiative excitation by the AGN may contribute additionally to exciting CO, a possibility that we have not included in our computations.

In summary, we infer densities of $\sim$$10^{3-4} {\rm \ cm^{-3}}$ for the Inner filament and densities $\sim$$10^{2-3} {\rm \ cm^{-3}}$ for the Western filament.\footnote{As an instructive comparison, densities of $\sim$$10^{2-3} {\rm \ cm^{-3}}$ are comparable with the bulk densities of Galactic GMCs, which have bulk temperatures of $\sim$10\,K.  Densities of $\gtrsim$$10^{4} {\rm \ cm^{-3}}$ are comparable with that in the central molecular zone of our Galaxy, spanning a radius of $\sim$200\,pc from the Galactic center, where the molecular gas has a bulk temperature of $\sim$50\,K \citep{oka07}.  Both \citet{hel97} and \citet{oka01} attribute the higher internal gas pressure of GMCs in the inner region of the Galaxy to higher intercloud gas pressures.}  The quoted ranges in densities depend upon the (unknown) gas temperature, which we assume to lie in the range $\sim$30--100\,K (with the density decreasing with increasing temperature).  In the computations of \citet{can16}, where the gas chemistry, ionization, and emission is solved self-consistently as a function of energetic particle density, they show that the gas remains primarily molecular only at temperatures below a few tens of K (see the left panels of their Figs.\,2 and 3); at such temperatures, the corresponding densities in the Inner and Western filaments are at the higher end of the ranges quoted above.  Over the range of temperatures considered, we find that the filaments have a thermal gas pressure, $n_e T \sim 10^{4.5-5.5} {\rm \ cm^{-3} \ K}$, that is 1--2 orders of magnitude lower than the surrounding X-ray gas pressure, $n_e T \sim 10^{6.5} {\rm \ cm^{-3} \ K}$.


\subsection{Gas emitting in HCN, HCO$^+$, and CN}\label{Gas HCN}

Using the IRAM 30-m telescope, \citet{sal08b} detected HCN(3-2) towards the center of \object{NGC 1275}.  Using the JCMT, \citet{bay11} also detected HCO$^+$(3-2), CN(2-1), and C$_2$H(3-2) towards the center of \object{NGC 1275}.  Furthermore, \citet{bay11} detected HCO$^+$(3-2) and, tentatively, CN(2-1) towards the Eastern filaments.  Both \citet{sal08b} and \citet{bay11} assumed that all these lines originate from molecular gas that is much denser than that traced in CO; in Galactic GMCs, these lines trace dense condensations (cores), whereas CO traces primarily the surrounding more diffuse gas.  As explained in $\S\ref{Electron-impact excitation}$, emission in the low rotational transitions of HCN, HCO$^+$, or CN should not be automatically assumed to originate from relatively dense gas ($\gtrsim 10^{5-6} {\rm cm^{-3}}$) without first considering the role electrons might play in exciting these molecules.  

As an illustrative example, we consider whether the observed HCN emission towards the center of \object{NGC 1275} can originate from the same gas as that emitting in CO, for which we find H$_2$ gas densities of at most $\sim$$10^{4} {\rm \ cm^{-3}}$ (see $\S\ref{Gas CO}$).  The latter is about 4 orders of magnitude below the critical density (for H$_2$-impact excitation) of HCN(3-2), so that in the gas traced by CO, HCN is only very weakly excited by collisions with H$_2$ (in addition to the weak excitation provided by radiation from the CMB, as well as self-absorption of photons produced by the emitting gas).  If the molecular gas has a sufficiently high ionization fraction, however, collisions with electrons can more strongly excite HCN.  If the electron density in the gas is of order $\sim$$0.1 {\rm \ cm^{-3}}$ as computed in $\S\ref{Electron-impact excitation}$ for secondary electrons released through collisional ionization, then for a H$_2$ gas density of $\sim$$10^3$--$10^4 {\rm \ cm^{-3}}$ as inferred for that traced in CO ($\S\ref{Gas CO}$), the corresponding ionization fraction is $\sim$$10^{-5}$--$10^{-4}$.  If X-ray irradiation contributes additionally to the release of electrons through photoionization, then the ionization fraction can be even higher.
We note that, according to the calculations of \citet{fer09}, the gas can reach an ionization fraction of up to nearly $10^{-3}$ and still remain primarily molecular.   

In Figure\,\ref{HCN model}, we plot the optical depths, excitation temperatures, and brightness temperatures of HCN(1-0), HCN(2-1), and HCN(3-2) as a function of $n(\rm H_2)$ at an illustrative gas temperature of $T_{\rm k} = 50$\,K, both for entirely neutral molecular gas (ionization fraction $x_{\rm e} = 0$; left column) and for molecular gas with an ionization fraction of $x_{\rm e} = 10^{-4}$ (right column).  We adopt in this figure $N_{\rm HCN}/\Delta{v} = 3 \times 10^{13} {\rm \ cm^{-2} \ km^{-1} \ s}$, permitting a direct comparison with Figure\,\ref{CO model 1} (middle panel) for CO(1-0), CO(2-1), and CO(3-2) given the same linewidth in HCN as in CO of $\sim$$100 {\rm \ km \ s^{-1}}$ and a fractional HCN abundance comparable to that of Galactic GMCs of $\rm [HCN/H_2] = 10^{-8}$ (3 orders of magnitude lower than the fractional CO abundance).  
The optical depths, excitation temperatures, and hence brightness temperatures of the HCN lines are significantly higher with (right column of Fig.\,\ref{HCN model}) than without (left column of Fig.\,\ref{HCN model}) collisional excitation by electrons.  The increase in brightness temperature of the HCN(3-2) line with increasing ionization fraction can be more easily seen in Figure\,\ref{HCN(3-2)} for $x_{\rm e} = 0, \ 10^{-4}, \rm \ and \ 10^{-3}$.  Like for CO, the change in optical depth with $n(\rm H_2)$ reflects the change in the relative populations of the different rotational $J$ levels as the particle densities change; note that, in the right column of Figure\,\ref{HCN model} where the ionization fraction is held fixed, the electron density increases in proportion with $n(\rm H_2)$.
With the addition of collisional excitation by electrons, the brightness temperatures of the HCN lines can be a significant fraction of the gas temperature even at H$_2$ gas densities far below the critical density for collisional excitation by H$_2$ molecules.  The latter is $n_{\rm crit} \sim 2 \times 10^6 {\rm \ cm^{-3}}$ for HCN(1-0), $n_{\rm crit} \sim 1 \times 10^7 {\rm \ cm^{-3}}$ for HCN(2-1), and $n_{\rm crit} \sim 7 \times 10^7 {\rm \ cm^{-3}}$ for HCN(3-2) at a gas temperature of 50\,K ($n_{\rm crit}$ decreasing slowly with increasing temperature).  

Although the addition of electron-impact excitation means that HCN can be reach significant excitation temperatures at H$_2$ gas densities far below the critical density for H$_2$-impact excitation, the decrease in optical depth as the particle (H$_2$ and electron) density increases means that the brightness temperature never reaches anywhere close to the gas temperature under the physical conditions considered in Figure\,\ref{HCN model}.  As reported by \citet{sal08b}, the peak brightness temperature of the HCN(3-2) line is a factor of $\sim$20 lower than that in CO(2-1) towards the center of NGC\,1275 (correcting for the different beam sizes in both these lines).  The widths of both lines are comparable.  For the same gas density as in CO of $n(\rm H_2) \sim 10^{4} {\rm \ cm^{-3}}$ and also same depth for the emitting gas of $\sim$10\,pc so that $N_{\rm HCN}/\Delta{v} = 3 \times 10^{13} {\rm \ cm^{-2} \ km^{-1} \ s}$, we find from Figure\,\ref{HCN model} (right column) a predicted peak brightness temperature of $T_{\rm B} \sim 0.6$\,K.  The latter is a factor of $\sim$40 below the predicted peak brightness temperature in CO(2-1) of $\sim$25\,K for H$_2$ gas having the same physical properties (see $\S\ref{Gas CO}$), in rough accord with the observed factor of $\sim$20 between the measured peak brightness temperatures of these two lines.  (We emphasize that our model predictions for the relative brightness temperatures in CO and HCN make no assumptions about the gas geometry or filling factor, but simply that both lines originate from the same gas with a uniform density and temperature.)  For an even higher ionization fraction but otherwise the same physical parameters for the molecular gas, the peak brightness temperature in HCN(3-2) would be even higher as shown in Figure\,\ref{HCN(3-2)}, bringing our model predictions to even closer accord with the measurements.  This simple computation demonstrates that HCN(3-2) can originate from the same gas that emits in CO(2-1) and CO(3-2) provided that the ionization fraction is $\sim$$10^{-4}$ or higher.

In the scenario where HCN is excited primarily by electrons, we cannot rule out the possibility that a portion of the molecular gas it traces has a higher density, but a smaller filling factor, than that traced by CO.  For example, if the HCN-emitting gas has a factor of $\sim$10 higher gas density of $n(\rm H_2) \sim 10^{5} {\rm \ cm^{-3}}$ than the CO-emitting gas but the same ionization fraction of $\sim$$10^{-4}$, for the same $N_{\rm HCN}/\Delta{v} = 3 \times 10^{13} {\rm \ cm^{-2} \ km^{-1} \ s}$ the depth of the HCN-emitting gas is a factor of $\sim$10 smaller ($\sim$1\,pc) and consequently also the filling factor ($\sim$0.001).  (The contribution of the denser gas to the overall CO emission is therefore negligible.)  For this denser component, the predicted peak brightness temperature in HCN(3-2) is $\sim$5\,K.  Including also the HCN(3-2) emission produced by the same (more diffuse) gas as that dominating the CO emission and observed over a beam that includes both the lower and higher density gas components, the predicted peak brightness temperature of the overall HCN(3-2) emission is now $\sim$1\,K, a factor of $\sim$25 below and in close accord with the observed factor of $\sim$20 below the predicted peak brightness temperature in CO(2-1).  
This simple computation demonstrates that, for an ionization fraction of $\sim$$10^{-4}$ or higher, in addition to a comparable contribution from the more diffuse but thicker column of gas with a higher filling factor that dominates the emission in CO(2-1) and CO(3-2), HCN(3-2) also can originate from gas having a higher density but a lower column and consequently also lower filling factor, giving results that are consistent with the observations.  Whether at predominantly a single density or at two different bulk densities (with that at the lower density dominating the emission in CO), or spanning a range of densities (with lower filling factors at increasing densities), the important point here is that the molecular gas emitting in HCN(3-2) can be orders of magnitude lower in density than the critical density for H$_2$-impact excitation of this line provided that its ionization fraction is $\sim$$10^{-4}$ or higher.  Thus, the HCN emission from \object{NGC 1275} (and, by extension, other cluster central elliptical galaxies) need not originate from very dense gas characteristic of dense condensations in Galactic GMCs where stars form.

\section{Conversion between CO Luminosity and Molecular Gas Mass}\label{CO conversion}
We review the physical basis for a relationship between CO luminosity and mass of molecular (primarily H$_2$) gas in our Galaxy, as well as in galaxies similar to the Milky Way (i.e., normal star-forming spiral galaxies).  Our intentions are to make explicit the conditions (often largely forgotten or ignored) under which this relationship can be used to derive molecular gas masses, so as to make clear whether this relationship can be applied to cluster central elliptical (and, indeed, any other types of) galaxies.

As first pointed out by \citet{lis82} and \citet{you82}, observations in our Galaxy of relatively dense (bulk densities of $\sim$$10^2$--$10^3 {\rm \ cm^{-3}}$) and primarily molecular gas clouds, as well as relatively diffuse (densities of $\sim$300--$500 {\rm \ cm^{-3}}$) and only partially molecular gas clouds, indicate a linear relationship between their measured CO intensity in the CO(1-0) line (expressed as an integrated brightness temperature), $W_{\rm CO}$, and inferred column density of H$_2$ gas, $n({\rm H_2})$, such that
\begin{equation}
N_{\rm H_2} \approx (2-4) \times 10^{20} \ {W_{\rm CO} \over (\rm K \ km \ s^{-1})} {\rm \ (cm^{-2})} \ \ , \label{Eq1}
\end{equation}
where the coefficient is widely referred to as the $X$ factor (i.e., $X \approx (2-4) \times 10^{20}$).  In clouds that are primarily molecular, this relationship has been found to hold from small dark clouds to GMCs (which span sizes from $\sim$5--100\,pc); in the case of GMCs, this relationship holds for the entire cloud as well as their denser inner regions more closely associated with star formation.  Because H$_2$ does not emit at temperatures typical of the aforementioned gas clouds, $n({\rm H_2})$ is determined indirectly through inferred dust masses and an assumed dust-to-gas mass ratio, or $\gamma$-ray emission arising from the interactions of cosmic rays with H$_2$.  
%
%
Because CO is (very) optically thick at the column densities of both dark clouds and GMCs, and can have (depending on the line-of-sight through the Galaxy) significant opacity (optical depth $\tau \approx 1$) even in diffuse partially-molecular clouds, the underlying reason for any (let alone a linear) relationship between $n({\rm H_2})$ and $W_{\rm CO}$ remains a subject of intensive debate.  

Two diametrically opposite views have emerged to explain the observed empirical relationship expressed in Eq.(\ref{Eq1}), each of which is tied to a different picture for the physical nature of GMCs in our Galaxy \citep[e.g., review by][and references therein]{mck07}.  We now examine each viewpoint in turn.

\subsection{Virialized Spherical Clouds}\label{spherical}
The traditional viewpoint advocates GMCs to be roughly spherical structures that are virialized; i.e., at equilibrium between internal pressure and self gravity, rather than being confined by the external pressure of a higher-temperature component of the ISM.  \citet{lar81} discovered that the velocity dispersion (as measured from the CO linewidth) of GMCs increases with size, implying that turbulence (i.e., non-thermal motion) ---  the nature of which remains poorly understood --- rather than thermal gas pressure is primarily responsible for supporting GMCs against their own gravity; in the largest molecular clouds with sizes of $\sim$100\,pc, support by non-thermal pressure is $\sim$50 times stronger than that provided by thermal pressure.\footnote{The line-of-sight (1 dimensional) root-mean-square velocity dispersion of CO, denoted by $\sigma_v$ as used throughout this manuscript, is $0.19 {\rm \ km \ s^{-1}}$ at 10\,K.  Note that \citet{lar81} defines $\sigma_s$ as the 3-dimensional root-mean-square velocity dispersion, whereby for isotropic turbulence $\sigma_s  = \sqrt{3} \sigma_v$.}  Implicit in this viewpoint, as reflected in the computations below, is that magnetic fields do not provide significant support.  Armed with this recognition, \citet{dic86} derived a relationship between the observed CO luminosity and total molecular cloud mass in external galaxies, and \citet{sol87} for the observed CO luminosity and mass of individual GMCs in our Galaxy.  Both proposed a similar underlying physical basis for this relationship.   

We briefly review the central arguments of \citet{dic86} and \citet{sol87} to make explicit the conditions under which Eq.(\ref{Eq1}) (which also can be written as a relationship between H$_2$ gas mass and CO luminosity; see below) can be used to infer the mass of H$_2$ gas in galaxies.  We begin with the arguments laid out by \citet{sol87} for GMCs in our Galaxy, and follow their notation in the equations below.   For a spherical gas cloud in virial equilibrium and assumed to have a power-law radial density profile, its mass, $M_{\rm VT}$ (where VT denotes the virial theorem), can be immediately derived from its velocity dispersion, $\sigma_v$, as measured from the observed width of the CO line emitted by this cloud, using the equation
\begin{equation}
M_{\rm VT} = 1052 \ \left({\sigma_v \over {\rm  km \ s^{-1}}}\right)^2 \ {R_e \over {\rm pc}} \ ({\rm M_{\sun}}) \ , \label{Eq2}
\end{equation}
where $R_e$ is the effective radius of the cloud such that $\pi {R_e}^2$ is the cloud area.  \citet{sol87} found that the mass of GMCs as derived from the virial theorem (Eq.(\ref{Eq2})) is correlated with the observed luminosity of these clouds as given by the empirical relationship
\begin{equation}
M_{\rm VT} = 39 \left({L_{\rm CO} \over {\rm K \ km \ s^{-1} \ pc^2}}\right)^{0.81 \pm 0.03} \ ({\rm M_{\sun}})  \label{Eq3}
\end{equation}
that holds over four decades in cloud masses from $\sim$$10^{3-7} {\rm \ M_\sun}$.  
The following is a synopsis of the arguments laid out by \citet{sol87} for the physical basis of the empirical relationship given by Eq.(\ref{Eq3}).

From their improved and more comprehensive observations of GMCs compared with those of \citet{lar81}, \citet{sol87} find that these clouds have velocity dispersions related to their mean geometrical size $S = D \ \rm tan (\sqrt{\sigma_l \ \sigma_b})$ (note that, defined in this manner, the geometrical shape of the cloud is rectangular), where $D$ is the distance to the cloud and $\sigma_l$ ($\sigma_b$) its angular extent in Galactic longitude (latitude), according to the empirical relationship
\begin{equation}
{\sigma_v} = 1.0 \pm 0.1 \ \left({S \over \rm pc}\right)^{0.50 \pm 0.05} {\rm \ (km \ s^{-1})} \  \label{Eq4}
\end{equation}
over the observed size range from a few parsecs to a few tens of parsecs.  Assuming no self-absorption within a cloud, its luminosity is simply the product of its (circular) projected surface area, $\pi {R_e}^2$, where $R_e$ is the effective radius of the cloud (for which \citet{sol87} find $R_e = (3.4 / \sqrt{\pi}) S$), and average surface brightness, which for a Gaussian line profile is given by $\sqrt{2 \pi} T_{\rm o} \sigma_v$, where $T_{\rm o}$ is the peak brightness temperature at the line center averaged over the cloud, so that
\begin{dmath}
L_{\rm CO} = 11.6 \sqrt{2 \pi} \ \left({T_{\rm o} \over {\rm K \ km \ s^{-1}}}\right) \\
\left({\sigma_v \over {\rm  km \ s^{-1}}}\right) \left({S \over {\rm pc}}\right)^2 \ (\rm K \ km \ s^{-1} \ pc^2) \ .  \label{Eq5}
\end{dmath}
To justify the assumption that the CO emission from a given surface of the cloud at a particular radial velocity, $v$, is not absorbed by intervening gas in the cloud, \citet{sol87} adopted what they coined the "mist" model for GMCs.  This model assumes that a GMC consists of a large number of small, optically-thick regions (droplets) that may overlap in projection but do not overlap in radial velocity along a given line of sight.  In this way, a GMC is effectively optically thin (i.e., photons emitted by an optically-thick droplet are not absorbed along the line of sight towards the observer) at each radial velocity, and therefore the CO brightness of the cloud traces its column density (i.e., the column density of the cloud is dominated by the droplets, with the diffuse gas between the droplets contributing negligibly to the column density).  Applying Eq.(\ref{Eq4}) in Eq.(\ref{Eq5}) yields
\begin{dmath}
L_{\rm CO} = 30 \left({T_{\rm o} \over {\rm K \ km \ s^{-1}}}\right) \left({\sigma_v \over {\rm  km \ s^{-1}}}\right)^5 \ (\rm K \ km \ s^{-1} \ pc^2) \ .  \label{Eq6}
\end{dmath}

Substituting into Eq.(\ref{Eq6}) the virial mass as expressed by Eq.(\ref{Eq3}) and utilizing the empirical relationship between size and velocity dispersion of GMCs as expressed by Eq.(\ref{Eq4}), 
\begin{dmath}
M_{\rm VT} = 132 \left[ \left({L_{\rm CO} \over {\rm K \ km \ s^{-1} \ pc^2}}\right) / \left({T_{\rm o} \over {\rm K \ km \ s^{-1}}}\right) \right]^{4/5} \ ({\rm M_{\sun}}) \ .  \label{Eq7}
\end{dmath}
\citet{sol87} found an average $T_{\rm o}$ of $\sim$4\,K for the GMCs they observed, thus yielding
\begin{equation}
M_{\rm VT} = 43 \left({L_{\rm CO} \over {\rm K \ km \ s^{-1} \ pc^2}}\right)^{4/5} \ ({\rm M_{\sun}}) \ , \label{Eq8}
\end{equation}
which is in good agreement with Eq.(\ref{Eq3}) and thus provides a physical basis for the empirical relationship between the virial masses of GMCs and their CO luminosities.

Rewriting Eq.(\ref{Eq5}) by noting that $\sqrt{2 \pi} T_{\rm o} \sigma_v = W_{\rm CO}$ gives
\begin{equation}
L_{\rm CO} = 11.6 \left({S \over {\rm pc}}\right)^2 W_{\rm CO} \ (\rm K \ km \ s^{-1} \ pc^2) \ .  \label{Eq9}
\end{equation}
The mass of a primarily H$_2$ gas cloud is related to its average column density $N({\rm H_2})$ by
\begin{dmath}
M = 11.6 \left({S \over {\rm pc}}\right)^2 \left(N({\rm H_2}) \over {\rm cm^{-2}} \right) \ \left(m_{\rm H_2} \over {\rm M_{\sun}} \right) \ ({\rm M_{\sun}}) \ ,  \label{Eq10}
\end{dmath}
where $m_{\rm H_2}$ is the mass of a hydrogen molecule.  Utilizing Eq.(\ref{Eq9}) and the empirical relationship between the virial mass and observed CO luminosity of GMCs as given by Eq.(\ref{Eq3}), Eq.(\ref{Eq10}) can be rewritten as
\begin{dmath}
N({\rm H_2}) = 6.1 \times 10^{21} \ \left(M \over {\rm M_{\sun}}\right)^{-1/4} W_{\rm CO} \ ({\rm cm^{-2}}) \ . \label{Eq11}
\end{dmath}
For (virial) masses $\sim$$5 \times 10^4$--$1 \times 10^6$, Eq.(\ref{Eq11}) gives $N({\rm H_2}) = 2$--$4 \times 10^{22} \ W_{\rm CO}$, thus explaining the empirical relationship between the measured CO intensity and inferred column density of H$_2$ gas as given by Eq.(\ref{Eq1}) for GMCs in our Galaxy.  In diffuse and only partially molecular clouds, \citet{lis10} argue that the higher (physical) temperature of the cloud (resulting in a higher $T_{\rm o}$ for the CO line) is counterbalanced by the lower abundance of CO in the cloud (CO being dissociated by interstellar UV radiation, which penetrates deeper into clouds with lower column densities) to result --- entirely fortuitously --- in the same empirical relationship between their column densities and CO luminosities.

In external (spiral) galaxies, \citet{dic86} suggest that observations in CO trace an ensemble of clouds similar to GMCs in our Galaxy.  They argue that along any single line-of-sight, individual GMCs are very unlikely to overlap in projection, let alone in radial velocity.  Thus, along any single line-of-sight, the CO emission is effectively optically thin; i.e., the droplets that represent individual optically-thick regions in a single GMC in the mist model of \citet{sol87} now represent individual optically-thick GMCs in an ensemble.  Assuming that each cloud is virialized, and utilizing the empirical relationship between size and velocity dispersion observed for GMCs in our Galaxy, \citet{dic86} derived a relationship between the total molecular mass and CO luminosity of external galaxies that, in the modern form, is exactly the same as that for individual GMCs in our Galaxy.

Thus, in the picture where molecular clouds are quasi-spherical virialized structures (having a power-law radial density profile) supported only by thermal pressure and turbulence, the use of either Eq.(\ref{Eq3}) or Eq.(\ref{Eq8}) to derive the H$_2$-gas masses of these clouds from their CO luminosities requires these clouds to: 1) exhibit a dependence in velocity dispersion with size in accord with Eq.(\ref{Eq4}) as found for GMCs in our Galaxy; 2) not overlap in radial velocity along a given line of sight \citep[the mist model of][]{sol87}; and 3) have a peak brightness temperature at line center averaged over the entire cloud of $\sim$4\,K as measured for GMCs in our Galaxy.  On the first point, there is no obvious reason to suppose that, in cluster central elliptical galaxies, the molecular gas (traced in CO) exhibits a dependence in velocity dispersion with size in accord with that found for GMCs in our Galaxy.  In disk galaxies, GMCs are found predominantly in spiral arms, where (in the traditional view) atomic gas is compressed by density waves to form molecular gas.  By contrast, in all but one cluster central elliptical galaxies so far imaged in CO \citep{sal06,lim08,sal08a,hol09,rus14,mcn14,van16}, the molecular gas does not reside in a rotating disk, but instead in filamentary structures.  
Possible sources of strong turbulence in the ISM of cluster central elliptical galaxies are likely to be very different from those known to operate in the disk of our Galaxy and other star-forming spiral galaxies, a point we shall return to in $\S\ref{filamentary}$.  On the third point, the characteristic temperature of $> 20$\,K that we infer for the molecular gas traced by CO close to the center of \object{NGC 1275} is significantly higher than the bulk temperature of $\sim$10\,K for Galactic GMCs.  Thus, at the very least, the relationship between H$_2$-gas mass and CO luminosity as given by Eq.(\ref{Eq3}) or Eq.(\ref{Eq8}) needs to be modified in accord with the (currently unknown) relationship between size and velocity dispersion of molecular gas clouds in cluster central elliptical galaxies, as well as the (currently unknown) peak brightness temperature at CO line center averaged over individual clouds in these galaxies.  The manner in which both these factors affect the relationship between H$_2$-gas masses and CO luminosities are almost universally ignored, with attention being focussed instead on possible differences in metallicity --- by applying a simple correction factor that reflects the relative abundance of He to H depending on the metallicity --- between the object being studied and our Galaxy.  
If magnetic fields are invoked to help support the molecular gas, as invoked by \citet{fab08} to stabilize the thread-like structures observed in the optical line-emitting filaments against gravitational collapse, then the relationship between H$_2$-gas mass and CO luminosity as given by Eq.(\ref{Eq3}) or Eq.(\ref{Eq8}) will require even more modifications before it can be applied to cluster central elliptical galaxies.

There is good reason to suspect that the bulk of molecular gas traced by CO in \object{NGC 1275}, and for that matter many other cluster central elliptical galaxies, does not reside in virialized structures.  Virtually all GMCs in our Galaxy are actively forming stars, whereas in \object{NGC 1275} there is no evidence for recent star formation in (the bulk of) the molecular gas traced by CO (see $\S\,\ref{Introduction}$).  If not virialized, then there is no physical basis for a relationship between H$_2$-gas mass and CO luminosity in the manner described above.  

\subsection{Turbulent Structures}\label{filamentary}
Rather than being virialized quasi-spherical clouds, the diametrically opposite viewpoint advocates that GMCs in our Galaxy (and, by extension, in the disks of external galaxies) are self-gravitating filamentary structures created by supersonic turbulence (driven on large scales) in a magnetized ISM.  This viewpoint has gained strong support through observations by {\it Herschel}, which have revealed the ubiquity of filamentary structures in GMCs \citep[e.g., see review by][]{and15}.  In this picture, the empirical dependence between size and velocity dispersion of Galactic GMCs (i.e., $\sigma_v \propto S^{1/2}$; Eq.(\ref{Eq4})) simply reflects the scaling properties of supersonic turbulence in the ISM.  

\citet{she11a,she11b} have made the most detailed magnetohydrodynamic simulations thus far of model molecular gas clouds (with different input metallicities) created by large-scale turbulence with resulting physical properties (i.e., bulk and globally-averaged column density, bulk temperature, as well as dependence between velocity dispersion and size) comparable with those of Galactic GMCs.  Their simulations ignore self gravity, and so their model filaments need not necessarily be self gravitating.  The overarching goal of \citet{she11a,she11b} is to understand why Galactic GMCs obey the empirical $X$ factor that relates their H$_2$ column density ($n({\rm H_2})$) to CO intensity ($W_{\rm CO}$) as expressed by Eq.(\ref{Eq1}), and also the factors that most influence the value of $X$.  In their simulations, \citet{she11a,she11b} included a background (i.e., interstellar) ultraviolet (UV) radiation field, as well as a treatment of chemistry to follow the formation and destruction of numerous chemical species including H$_2$ and CO.  H$_2$ forms on (and is subsequently ejected from) the surfaces of dust grains, whereas CO forms in the gas phase as a product of ion-neutral chemistry.  Crucially, while absorption of (interstellar) UV radiation by dust helps shields both molecules from photodissocation, H$_2$ can, in addition, protect itself from UV radiation through self-shielding (electronic excitation in the Lyman and Werner bands that have a higher probability of decaying to a lower-excitation rather than a dissociative state) even in gas with relatively low column densities, whereas CO is not able to self shield.  As a consequence, the [CO/H$_2$] abundance can vary dramatically in a cloud depending on how deep UV penetrates from a given direction.  
Furthermore, while the mass-weighted average temperature ($\sim$20\,K) of the simulated molecular clouds is roughly comparable with that of Galactic GMCs, the temperature varies significantly throughout the cloud (for comparison, the volume-weighted average temperature is $\sim$50\,K) depending on the magnitude of shock heating, UV penetration, radiative cooling, and chemistry within a given region.  The physical properties of the model molecular gas clouds are therefore considerably more complex than those that might be naively envisioned for --- and perhaps more reflective of the true conditions within --- Galactic GMCs.
  
\citet{she11a,she11b} find that the $X$ factor varies for different lines of sight into their model clouds: along any single line-of-sight, the CO radiation that escapes from the cloud originates from a range of depths having different densities, temperatures, and [CO/H$_2$] abundance.  The radiation from any given depth can be partially if not entirely absorbed by intervening gas in the cloud; i.e., the mist model of \citet{sol87} in no way captures the complexity intrinsic to line radiative transfer from a turbulent medium.  As a consequence, there is no direct relationship between $n({\rm H_2})$ and $W_{\rm CO}$ as expressed by Eq.(\ref{Eq1}) for any line of sight into the model cloud.  Nevertheless, for the model cloud as a whole, \citet{she11a,she11b} find that the average $X$ factor lies in the range of values found empirically for GMCs in our Galaxy, reflecting a dependence between $W_{\rm CO}$ and the globally-averaged column density $\overline{N_{\rm H_2}}$.  The nearly constant $X$ factor observed for Galactic GMCs is therefore simply a reflection of nothing more than the limited range in physical properties exhibited by these clouds.  

Interestingly, \citet{she11a,she11b} find that, for clouds with physical properties otherwise similar to Galactic GMCs, the $X$ factor does not depend on the existence of a scaling relationship between size and velocity dispersion (e.g., Eq.(\ref{Eq4})), but only on the overall range in velocity exhibited by the gas cloud.  As mentioned above, any given line of sight into the cloud need not be effectively optically thin, but instead is usually optically thick: photons produced by gas at a given radial velocity have a higher probability of escaping from clouds having a larger velocity dispersion, resulting in a larger $W_{\rm CO}$ for clouds with otherwise similar physical properties.  
As a consequence, the $X$ factor scales roughly as $X \propto \sigma_v^{-1/2}$.  The $X$ factor also depends on the bulk temperature of the cloud, scaling roughly as $X \propto T^{-1/2}$; as noted by \citet{she11b}, this relationship was derived assuming no changes in the cloud chemistry with bulk temperature, which is unlikely to be correct and hence must be treated with caution.  \citet{she11a}, however, was not able to reproduce the observed $X$ factor for diffuse and only partially molecular clouds in our Galaxy, finding a value for the $X$ factor that is four orders of magnitude higher than is inferred for these clouds.  The difference reflects the much lower [CO/H$_2$] abundance \citet{she11a} find in their simulations compared with the value inferred by \cite{lis10} for these clouds; in the simulations of \citet{she11a}, CO is effectively photodissociated by interstellar UV at the low column densities of diffuse molecular clouds, thus resulting in a very low [CO/H$_2$] abundance.  This finding also makes clear that, given the complex manner in which the shielding and chemistry of different molecular species vary with metallicity, it is far from obvious that the relationship between $L_{\rm CO}$ and $M({\rm H_2})$ for our Galaxy can be used for another galaxy with a different metallicity by simply applying a multiplicative factor related to its different metallicity, a formula that is commonly adopted in the literature.  

As emphasized by \citet{she11a,she11b}, the nearly constant conversion factor inferred for Galactic GMCs simply reflects the inherently limited range in the physical properties of these clouds.  Clouds with significantly different physical properties than Galactic GMCs 
are unlikely to have similar conversion factors.  
Even if we assume the same density, temperature, and metallicity, the [CO/H$_2$] abundance may be very different in cluster central elliptical galaxies compared with Galactic GMCs given the very different environments to which the molecular gas in these galaxies are exposed.  The simulations performed by \citet{she11a,she11b}, with the goal of replicating Galactic GMCs, show that [CO/H$_2$] can vary dramatically with spatial location in the molecular gas depending on the incident UV radiation field (which dissociates CO) and the dust content (which absorbs UV thus shielding CO from photodissociation).  Ultraviolet continuum and line emissions have been detected towards the core of the Perseus cluster \citep{dix96,bre06}, including the O~VI lines at 1032~\AA\ and 1035~\AA\ that \citet{bre06} attributed to gas cooling through $\sim$$10^{5.5} {\rm \ K}$ at a rate consistent with current upper limits on an X-ray cooling flow in the Perseus cluster.  The role of soft X-rays (which can penetrate further than UV, and dissociate both CO and H$_2$), which is emitted by one of the gas phases of the ISM, as well as the surrounding ICM, on the chemistry of the molecular gas has not been explored in the context of central cluster elliptical galaxies.  The infiltration of ICM X-ray gas into the molecular gas can reduce, through grain sputtering, the dust content, which in turn affects the H$_2$ formation rate and thus the [CO/H$_2$] abundance.  As shown by \citet{bay10}, highly energetic particles can have a significant effect on the chemistry of the molecular gas, altering the abundance of various molecules (although apparently not CO).  The gas chemistry also depends on the gas temperature; in $\S$\ref{Gas CO}, we showed that the molecular gas traced by CO in \object{NGC 1275} can have a higher characteristic temperature than that of Galactic GMCs. 
As mentioned above, even for the same gas chemistry, the $X$ factor depends on the gas temperature.  Finally, although immune to any scaling between size and velocity dispersion of the gas structures, the $X$ factor does depend on the gas velocity dispersion, which may have both larger values and exhibit a much larger range of values in cluster central elliptical galaxies than the smaller values and more limited range of values exhibited by Galactic GMCs.


\section{Nature of Molecular Gas}\label{nature molecular gas}

\subsection{Turbulence-Driven Ensembles?}
Motivated by the turbulence-driven nature of Galactic GMCs ($\S\ref{filamentary}$) together with the apparent lack of recent star formation in the (bulk of) molecular gas traced by CO in \object{NGC 1275}, here we propose a very different picture for the nature of molecular gas (and, by extension, many of the other ISM components) in cluster central elliptical galaxies than either the idealized picture of quasi-spherical gas clouds or self-gravitating turbulent and magnetized structures.  Our proposal makes no attempt to explain the morphology of the ISM (i.e., large-scale filamentary structures in the case of \object{NGC 1275}), but rather the nature of the gas components contained within.  

We speculate that, in the filaments of \object{NGC 1275}, (internal) turbulence is responsible for concentrating gas and dust in overdense regions.  (In this picture, turbulence can play either a constructive or destructive role, alternatively concentrating material in or dispersing material from overdense regions.)  In these overdense regions, the abundance of molecular gas is enhanced by in situ production and its destruction mitigated by both dust and self shielding.  The gas composition (i.e., molecular vs atomic$+$ionized) in these regions is controlled primarily by the density of highly-energetic particles (X-ray gas that infiltrates into the filaments), which can vary from one location to the next depending on the penetrating efficiency of these particles.  In more rarefied regions, in addition to dissociation by highly-energetic particles, molecular gas can be strongly dissociated also by soft X-ray and UV radiation, all of which help determine the gas composition in these regions (presumably more atomic$+$ionized than molecular).  The molecular gas ensembles thus created may be transient, pressure-confined features rather than virialized structures, thus explaining the apparent lack of recent star formation within the vast majority, if not all, of the CO filaments so far detected in \object{NGC 1275}.  

In the picture we propose, rather than being concentrated in GMCs analogous to those in our Galaxy, the molecular gas is distributed in complex patterns --- interspersed between atomic/ionized gas that occupies a larger volume of the ISM --- reflecting the nature of turbulence in the ISM, and the in-situ production versus destruction of molecular gas by the harsh radiation and particle field impinging from the surrounding ICM.  What then is the source of turbulence in the ISM of \object{NGC 1275}?  In our Galaxy, large-scale turbulence in the ISM is believed to be driven by energy input from, primarily, supernovae and magnetorotational instability \citep{mac04}, with instabilities in spiral shocks making a significant contribution in spiral arms \citep{kim06}.  In \object{NGC 1275}, the ISM turbulence may be driven by organized or turbulent motions in the surrounding ICM, which have been found to show a line-of-sight velocity dispersion of $160 \pm 10 {\rm \ km \ s^{-1}}$ in the cluster core \citep[in a region spanning a projected distance of 30--60\,kpc from the center;][]{hit16}.   Turbulence is likely generated by the AGN in \object{NGC 1275} through the inflation of bubbles in the ICM filled with relativistic plasma; these bubbles are visible as X-ray-evacuated bubbles, and sometimes also emit detectable synchrotron radio emission \citep{boh93,fab06}.  Once they reach a sufficiently high pressure to become buoyant, the bubbles rise through and drive turbulence in the ICM \citep[e.g.,][]{chu00,zhu14,zhu16}.
A U-shaped optical line-emitting filament known as the horseshoe filament, located behind (radially inwards of) a detached X-ray bubble, is interpreted as a filament caught up in a vortex generated behind this buoyantly rising bubble \citep{fab03}.  The existence of relatively strong turbulence in the ISM, driven by turbulence in the ICM, would explain the large CO linewidths of $\sim$50--$100 {\rm \ km \ s^{-1}}$ as measured at full-width half-maximum (FWHM) that we see even at the highest spatial resolutions so far achieved of $\sim$1\,kpc \citep[][this paper]{hol09,sal08a}, as well as the large H$\alpha$ FWHM linewidths of typically 50--$160 {\rm \ km \ s^{-1}}$ observed towards the filaments in \object{NGC 1275} at a spatial resolution of better than $\sim$300~pc \citep{hat06}.  \citet{zhu14} have shown that heating from the dissipation of this turbulence is able to counterbalance the radiative energy loss of the ICM.  Alternatively, or in addition, turbulence in the filament may be driven by magnetic instabilities, or by the influx of energetic particles into the filaments as suggested by \citet{can16}.

\subsection{Stability of Filaments}\label{stability filaments}
As previous emphasized by \citet{hol09}, using the Galactic conversion $M_{\rm H_2}/L_{\rm CO} = 4.6 \ M_\sun {\rm \ K \ km \ s^{-1} \ pc^2}$ in the CO(1-0) line \citep{sol87}, all the filaments so far imaged in CO have masses that far exceed their Jeans mass.  Computed in this manner, the Inner filament alone, which has a projected length of $\sim$2.6\,kpc, has a mass in molecular gas of $2 \times 10^9 {\rm \ M_\sun}$, comparable to that in the entire Galaxy.  Applying the Kennicutt-Schmidt law that relates the molecular gas surface mass density in spiral and starburst galaxies to their star formation rate surface density, we find a lower limit on the predicted star formation rate surface density for the filaments imaged in CO (lower limit as these filaments are not resolved in width) of $\sim$0.1--$10 {\rm \ M_{\odot} \ yr^{-1} \ kpc^{-2}}$; i.e., according to the Kennicutt-Schmidt law, each square kpc of these filaments should be exhibiting a star formation rate comparable with that of our entire Galaxy.  Yet, the filaments imaged in CO do not have spectra resembling HII regions \citep[except at an isolated location in the Eastern filament;][]{shi90}.

Here, we reconsider the stability of the Inner and Western filaments if they comprise (a bundle of) thin cylindrical filaments, as suggested by observations with the Hubble Space Telescope for the outer filaments \citep{fab08}.  For a cylindrical filament supported only by thermal gas pressure, the critical mass per unit length, $M_{\rm line, crit}$, above which the filament collapses due to its own gravity is $M_{\rm line, crit} \approx 16 \ M_\sun \ {\rm pc^{-1}} \times (T_{\rm k}/10\,{\rm K})$ \citep{inu97}; this condition holds irrespective of the external gas pressure \citep{fis12}.  Applying the Galactic $M_{\rm H_2}$--$L_{\rm CO}$ conversion and assuming for simplicity $\rm CO(2-1)/CO(1-0) = 1$ or $\rm CO(3-2)/CO(1-0) = 1$ gives $M_{\rm H_2} \sim 10^7$--$10^8 \ M_\sun$ per synthesized beam of diameter $\sim$1\,kpc along the Inner and Western filaments.  The corresponding mass per unit length, $M_{line}$, is therefore $\sim$$10^{4-5} \ M_\sun \ {\rm pc^{-1}}$.  Thus, $M_{line} \sim 10^{3-4} \ M_{\rm line, crit}$, implying that the Inner and Western filaments should collapse if supported by thermal gas pressure alone (even if $T_{\rm k} \gg 10$\,K).  To not collapse due to their own self-gravity, non-thermal pressure support (combination or turbulence and magnetic fields) would then have to be $\sim$$10^{3-4}$ higher than the thermal gas pressure.

Given the uncertainty in applying the Galactic $M_{\rm H_2}$--$L_{\rm CO}$ conversion to the molecular filaments in \object{NGC 1275} as explained above, it is more instructive to express the mass per unit length of a cylindrical filament in terms of its gas density, $n{\rm(H_2)}$, and cross-section radius, $R$, whereby $M_{line} = 0.16 \ n{\rm(H_2)} {(\rm cm^{-3})} \ R^2 {(\rm pc)} \ M_\sun {\rm \ pc^{-1}} = n{\rm(H_2)} {\rm (cm^{-3})} \ R^2 (\rm pc) \ M_{\rm line, crit} / [100 (T_{\rm k}/10\,{\rm K})]$.  For $n{\rm(H_2)} \sim 10^{2-4} {\rm \ cm^{-3}}$, \\
$M_{line} \sim 1$--$100 \ R^2 (\rm pc) \ M_{\rm line, crit} / (T_{\rm k}/10\,{\rm K})$, so that at $T_{\rm k} \sim 30$\,K the filament must have $R < 0.2$--2\,pc (radius decreasing with increasing density) so as not to collapse due to its own self-gravity.  As an instructive comparison, such cross-sectional radii are comparable to the radii of molecular clumps -- sites for the formation of star clusters -- in GMCs.

As mentioned in $\S\ref{Gas CO}$, the molecular gas traced in CO (as well as in HCN, HCO$^+$, and CN) has a thermal pressure about 1--2 orders of magnitude lower than that of the surrounding X-ray gas.  In the absence of other means of support, the molecular component of the filaments will be squeezed into thin threads until -- provided they do not collapse due to their own self-gravity -- their thermal pressure balances the external pressure and their internal gravity.  Whether this component undergoes gravitational collapses depends on whether their cross-sectional radii are larger that the upper limits computed above (assuming support by thermal gas pressure alone) of $\sim$0.2--2\,pc.  By comparison, \citet{fab08} show examples of H$\alpha$ filaments in NGC\,1275 having cross-sectional profiles larger than the stellar point-spread-function, which has a radius of $\sim$35\,pc at full-width half-maximum (the cross-sectional radius of the molecular component is not known).  The filaments they examined are well away from the center; towards the center, the presence of overlapping filaments makes the examination of individual filaments difficult.

Above, we presented a picture in which the gas components in the filaments are inherently turbulent.  If turbulence rather than thermal gas pressure dominates, the permissible upper limit in the cross-sectional radius, $R$, increases dramatically, scaling as the total velocity dispersion, $\sqrt{\sigma_{\rm therm}^2 + \sigma_{\rm turb}^2}$, where $\sigma_{\rm therm}$ is the thermal and $\sigma_{\rm turb}$ the turbulent velocity dispersion \citep{fie00}.  \citet{can16} find that, to produce the observed line ratio in [O I] $\lambda \, 63 \, \mu {\rm m}$ to [C II] $\lambda \, 157 \, \mu {\rm m}$, they need to invoke a turbulent velocity dispersion of 2--$10 {\rm \ km \ s^{-1}}$.  As an illustration here, for $\sigma_{\rm turb}$$\sim$$5 {\rm \ km \ s^{-1}}$ (about an order of magnitude higher than the thermal gas pressure at 30--100\,K), a cylindrical filament filled with gas at a density of $\sim$$10^{2-4} {\rm \ cm^{-3}}$ and temperature $T_{\rm k}$$\sim$30\,K will not collapse if its cross-sectional radius $R < 2$--20\,pc (radius decreasing with increasing density).  This range in cross-sectional radii is comparable with that exhibited by GMCs in our Galaxy.

\citet{fab08} argue that the optical emission-line filaments in NGC\,1275 are stabilized against collapse by magnetic fields.  For magnetized filaments, $M_{\rm line, crit}^{mag} = M_{\rm line, crit} \times (1 - \mathcal{M}/ |\mathcal{W}|)^{-1}$, where $\mathcal{M}$ is the magnetic energy density per unit length and $\mathcal{W} = -M_{\rm line}^2 G$ is the gravitational energy per unit length \citep{fie00}.  Assuming that the CO filaments have $M_{\rm line} = M_{\rm line, crit}$, for $R$$\sim$1\,pc (approximate upper limit as computed above for filaments supported only by thermal gas pressure, beyond which they will collapse), $B$$\sim$$7 {\rm \ \mu G}$.  Interestingly, the latter is roughly comparable with the value computed assuming a magnetic pressure equal to that of the surrounding X-ray gas pressure, giving $B$$\sim$$24 {\rm \ \mu G}$ \citep[e.g.,][]{fab08}.  Of course, if supported also by turbulent pressure, the magnetic field strength can be lower for the same maximum cross-sectional radius above which collapse is inevitable.

\section{Summary and Conclusions}\label{Summary}
We present observations in CO(3-2) towards the center of \object{NGC 1275}, the central giant elliptical galaxy of the Perseus cluster.  Over a central radius of $\sim$18\arcsec\ ($\sim$6.5\,kpc), we find that the overall spatial-kinematic distribution in CO(3-2) resembles that in CO(2-1) as previously mapped by \citet{lim08} and \citet{hol09}.  A comparison of the spectral profiles in CO(3-2) and CO(2-1) reveals:

\begin{itemize}

\item[1.]  towards the nucleus (within an angular distance of $\sim$0.5\,kpc), a brighter component having a similar spectral profile in both CO(2-1) and CO(3-2), together with a dimmer blueshifted component that extends to much higher blueshifted velocities in CO(3-2) than in CO(2-1).  The ratio in brightness temperature of CO(3-2) to CO(2-1) of the brighter component is $\sim$0.9.

\item[2.]  away from the nucleus (out to an angular distance of $\sim$6\,kpc), similar spectral profiles in both CO(2-1) and CO(3-2), and a ratio in brightness temperature between CO(3-2) and CO(2-1) of $\sim$0.4--0.5.

\end{itemize}

Apart from CO(2-1) and CO(3-2), HCN(3-2) also has been detected towards the center of \object{NGC 1275} \citep{sal08b}, and both HCO$^+$(3-2) and CN(2-1) detected towards and away from the center \citep{bay11}.  The critical densities for H$_2$-impact excitation of HCN(3-2), HCO$^+$(3-2), and CN(2-1) are 2--3 orders of magnitude higher than those for CO(2-1) and CO(3-2), suggesting the presence of H$_2$ gas much denser than that traced by CO.
Because collisional excitation by electrons has been suggested to play an important role in exciting atomic and ionic emission lines from \object{NGC 1275} in the optical and near-IR, as well as the ro-vibrational lines of H$_2$ in the near- and mid-IR, we consider whether collisional excitation by electrons also can contribute significantly to exciting the rotational transitions of trace molecules at millimeter and submillimeter wavelengths.  We find that:

\begin{itemize}

\item[3.]  for electrons to compete with H$_2$ molecules in collisionally exciting CO, the required ionization fraction is $\sim$$10^{-2}$.  \citet{fer09} show that such high ionization fractions, reached through collisional ionization by energetic particles, leaves the gas primarily atomic rather than molecular.  Thus, CO emission is likely to arise from collisional excitation by H$_2$ molecules rather than electrons.

\item[4.] for electrons to compete with H$_2$ molecules in collisionally exciting trace molecules with larger dipole moments such as HCN, HCO$^+$, and CN, the required ionization fraction is $\sim$$10^{-6}$--$10^{-5}$ (computed specifically for HCN).  Both \citet{fer09} and \citet{can16} show that such ionization fractions, reached through collisional ionization by energetic particles, can leave the gas primarily molecular.  Indeed, ionization fractions of $\sim$$10^{-4}$ are found in diffuse partially-molecular clouds in our Galaxy, where HCN(1-0) and HCO$^+$(1-0) emissions are produced through electron-impact excitation \citep{lis12}.  Thus, collisional excitation by electrons in relatively diffuse molecular gas must be considered before assigning emission in HCN, HCO$^+$, and CN to relatively dense molecular gas.

\end{itemize}

Given that CO is excited primarily by collisions with H$_2$ molecules, from the observed ratio in brightness temperature between the CO(3-2) and CO(2-1) lines, assuming a temperature in the range $\sim$30--100\,K,  cylindrical filaments seen from the side, and a relative CO abundance of $\rm [CO/H_2] \sim 10^{-5}$ as in Galactic GMCs, we find that:

\begin{itemize}

\item[5.] towards the nucleus, the brighter component in the spectral profile traces H$_2$ gas with a density of $\sim$$10^3$--$10^4 {\rm \ cm^{-3}}$ and a depth for the emitting column along the line of sight of $\sim$10\,pc.


\item[6.] away from the nucleus, the CO emission traces H$_2$ gas with a density of $\sim$$10^2$--$10^3 {\rm \ cm^{-3}}$ and a depth along the line of sight of $\sim$100\,pc.

\end{itemize}

By including electron-impact excitation, we considered whether the HCN(3-2) emission observed towards the center of \object{NGC 1275} can be produced by the same gas as that emitting in CO(2-1) and CO(3-2).  Assuming a relative abundance in HCN comparable with that in Galactic GMCs of $\rm [HCN/H_2] \sim 10^{-8}$, we find that:

\begin{itemize}

\item[7.]  the relative brightness of the HCN(3-2) and CO(2-1) lines towards the nucleus can be produced self-consistently in molecular gas having a density of $\sim$$10^4 {\rm \ cm^{-3}}$ provided that the ionization fraction is $\sim$$10^{-4}$. 

\end{itemize}

\noindent The detection of HCN(3-2), as well as HCO$^+$(3-2) and CN(2-1), therefore does not necessarily imply high-density gas analogous to dense cores in Galactic GMCs where stars form.  Instead, in \object{NGC 1275}, these lines can be produced by the same molecular gas as that traced in CO, but with ionization fractions many orders of magnitude higher than that of Galactic GMCs.

We examined whether the conversion between CO luminosity and H$_2$-gas mass inferred for our Galaxy, and widely used in other galaxies, is likely applicable in cluster central elliptical galaxies.  In the traditional viewpoint whereby Galactic GMCs are assumed to be spherical virialized clouds supported primarily by turbulence, the physical basis for a relationship between CO luminosity and H$_2$-gas mass makes use of: 1) a dependence in velocity dispersion with size in accord with Eq.(\ref{Eq4}) as found for Galactic GMCs; 2) no overlap in radial velocity along a given line of sight; and 3) a peak brightness temperature at line center averaged over the entire cloud of $\sim$4\,K as measured for Galactic GMCs.  Although work using the Galactic CO luminosity and H$_2$-gas mass conversion usually only remark on differences in metallicity by applying a simple correction (multiplicative) factor (an approach that is likely to be far too naive given the effect of metallicity on shielding and chemistry of CO), differences in the other factors as listed above --- which are often ignored --- must also be considered.  Given the likely very different sources of turbulence in cluster central elliptical galaxies compared with our own Galaxy (or other normal star-forming disk galaxies), the molecular gas in these galaxies is unlikely to have the same dependence in velocity dispersion with size as is found for Galactic GMCs.  Furthermore, given the very different environment in which the molecular gas is immersed (irradiation by the surrounding X-ray gas, infiltration of the X-ray gas into the molecular gas), this gas may be at a significantly higher bulk temperature and its chemistry entirely different.

Observations over the past several years reveal that, rather than comprising virialized quasi-spherical clouds, Galactic GMCs are self-gravitating filamentary structures.  Theoretical models suggest that GCMs are created by supersonic turbulence driven on large scales in a magnetized ISM.   Studies by \citet{she11a,she11b} find that the Galactic conversion between CO luminosity and H$_2$-gas mass is merely a statistical relationship averaged over an entire cloud having the narrow range in physical properties displayed by Galactic GMCs.  Molecular gas ensembles having different physical properties, as is likely to be the case in cluster central elliptical galaxies, need not have the same, if any singular, relationship between CO luminosity and H$_2$-gas mass.  Rather than relying on a conversion between CO luminosity to H$_2$-gas mass (that may be pure guesswork), we stress instead that line ratios in a given molecular species, along with a proper understanding of the relevant excitation mechanisms, provide a more reliable method to infer the physical properties of the molecular gas in cluster central elliptical galaxies.  In this way, we find that:

\begin{itemize}

\item[8.] if supported by thermal pressure alone at densities $\sim$$10^{2-4} {\rm \ cm^{-3}}$ and temperatures $\sim$30\,K, to not collapse due to their own self-gravity the molecular component(s) of the filaments must (each) have a cross-sectional radius no larger than $\sim$0.2--2\,pc (decreasing with increasing density).  Additional support provided by turbulence or magnetic fields increases this maximum permissible cross-sectional radius above which collapse is inevitable.

\end{itemize}

We propose that in the ISM of cluster central elliptical galaxies, turbulence may be primarily responsible for concentrating dust and molecular gas in overdense regions, where the abundance of molecular gas is further enhanced by in situ production and its destruction reduced by both dust and self shielding. In these regions, the gas composition (i.e., molecular vs atomic+ionized) is controlled primarily by the density of energetic particles (X-ray gas that penetrates into the atomic and molecular ISM), which can vary from one location to the next depending on the penetrating efficiency of these particles. In more rarefied regions, the molecular gas can be strongly dissociated by soft X-ray and UV radiation, as well as by energetic particles, all of which help determine the gas composition in these regions (presumably more atomic+ionized than molecular).  The molecular gas need not be concentrated in GMCs analogous to those in our Galaxy, but distributed in complex structures --- interspersed between atomic/ionized gas that occupy a larger volume of the ISM --- reflecting the nature of turbulence in the ISM, and the in-situ production versus destruction of molecules by the harsh radiation and particle field impinging from the surrounding ICM.  In this picture, rather than being virialized structures analogous to GMCs, the molecular gas resides in transient and pressure-confined structures, thus explaining the apparent lack of recent star formation within the vast majority if not all the CO filaments so far detected in NGC 1275.

\acknowledgments

\acknowledgements  Work by J. Lim on this project is supported by the Research Grants Council of Hong Kong through grants HKU\,704011P and 17303414.  Dinh-V-Trung acknowledges the financial support of the National Foundation for Science and Technology Development (Nafosted) under grant number 103.99-2014.82.  We are grateful to R. Jay Gabany for providing us the beautiful and inspiring picture of the Perseus cluster shown in Figure\,\ref{NGC 1275}.  Finally, we thank M. Gurwell for helping us reduce the CO(3-2) data.

{\it Facilities:} \facility{SMA}.




\clearpage



\renewcommand{\baselinestretch}{1.0}

\onecolumn
\begin{figure}
\center
\vspace{-4cm}
\epsscale{1.0}
\plotone{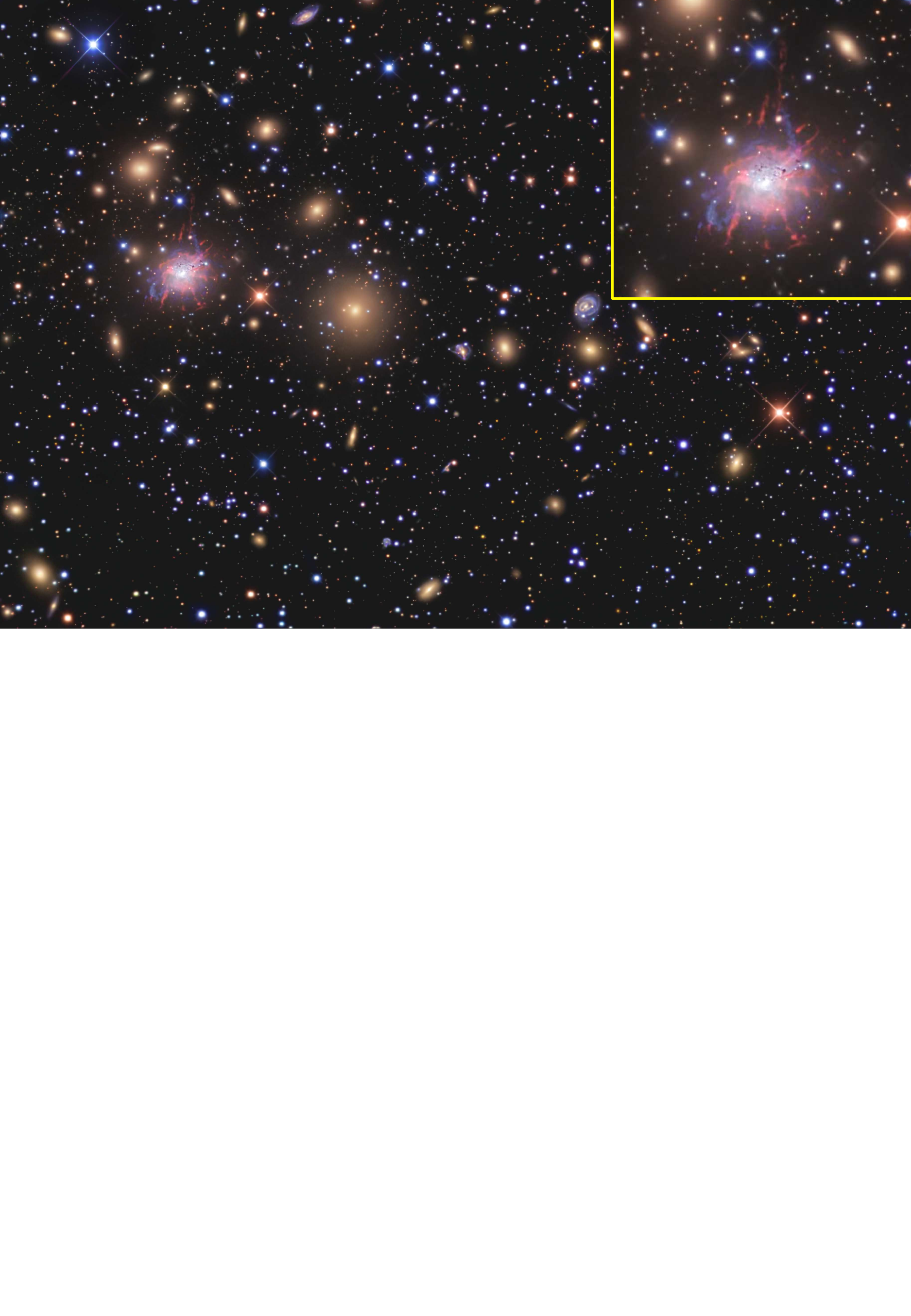}
\vspace{-12.5cm}
\renewcommand{\baselinestretch}{1.0}
\caption{A color composite of the Perseus cluster taken in broadband Red, Green, Blue, and Clear Luminance filters, as well as in a narrowband H$\alpha$ filter (colored red), at the 0.5-m telescope of the BlackBird Observatory (courtesy R. Jay Gabany).  The image spans $20\farcm4 \times 30\farcm6$.  NGC\,1275, the central giant elliptical galaxy of the Perseus cluster, is shown magnified in the inset at the upper right.  It hosts blue light from relatively young stars and a filamentary H$\alpha$ nebula, visible only in this but not other cluster elliptical galaxies contained in this image.}
\label{NGC 1275}
\end{figure}
\clearpage

\begin{figure}
\center
\vspace{-4cm}
\epsscale{0.95}
\plotone{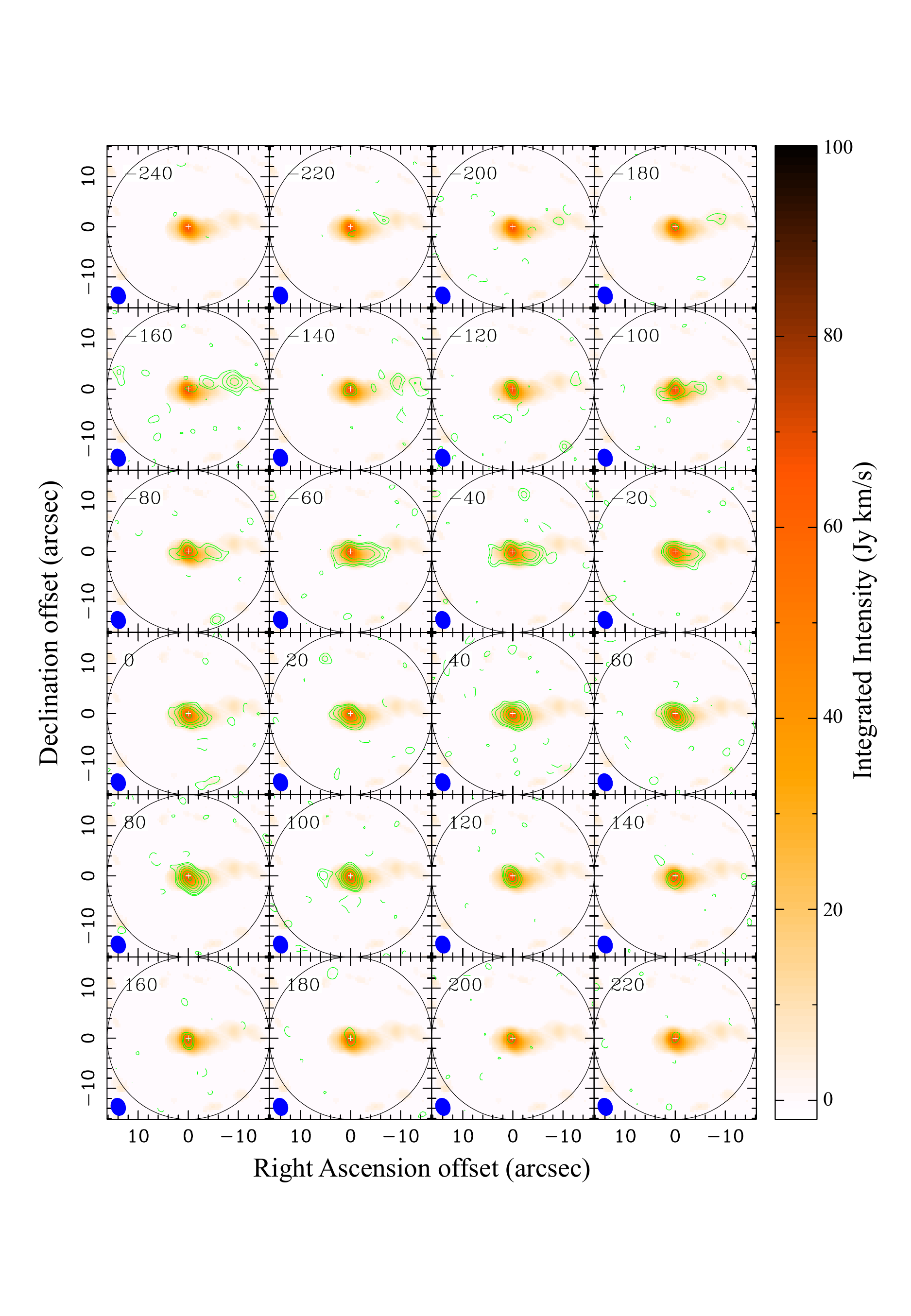}
\vspace{-2.6cm}
\renewcommand{\baselinestretch}{1.0}
\caption{Channel maps of the CO(3-2) emission from NGC\,1275.  Contours show the intensity of the line emission averaged over a velocity width of $20 {\rm \ km \ s^{-1}}$ centred at a velocity (measured, in units of ${\rm km \ s^{-1}}$, with respect to the systemic velocity of NGC\,1275; see text) as indicated in the individual panels.  Contour levels are plotted at ($-2$, 2, 3, 5, 7, 10, 13, 16, and $19) \times \sigma$, where $\sigma = 21 {\rm \ mJy \ beam^{-1}}$ is the rms noise fluctuation.  Colors show the line intensity integrated through the velocity range over which CO(3-2) is detectable (moment 0), constructed in the manner as explained in the text ($\S\ref{structure}$).  The integrated intensity levels are indicated by the color bar on the right.  All the images in CO(3-2) shown in this and the following figures have been convolved to give a synthesized beam ($3\farcs43 \times 2\farcs75$ and a position angle of $19\degr.9$), as indicated by the ellipse at the lower left corner, that is the same size as that obtained in observations in CO(2-1) by \citet{hol09} to permit computations of line ratios.  The cross indicates the position of the continuum emission from the AGN.  The large black circle in each panel indicates the primary beam of the SMA antennas.}
\label{CO3-2_channelmaps_mom0_col}
\end{figure}
\clearpage

\begin{figure}
\center
\vspace{-4cm}
\epsscale{1.0}
\plotone{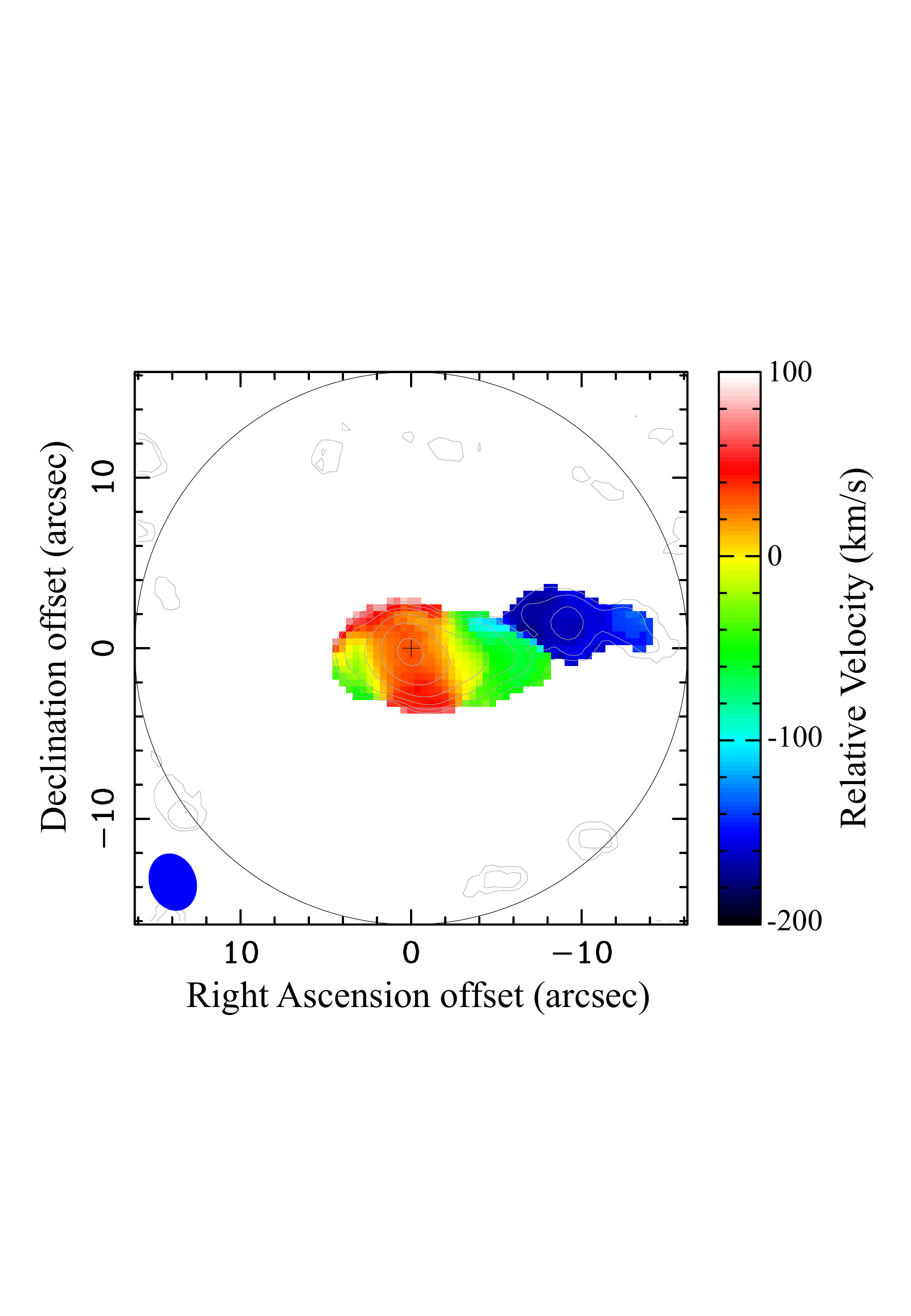}
\vspace{-5.5cm}
\renewcommand{\baselinestretch}{1.0}
\caption{Colors show the intensity-weighted mean velocity (moment 1) of the CO(3-2) emission with values as indicated by the color bar on the right.  Contours show the integrated line intensity (moment 0), identical to that shown in color in Fig\,\ref{CO3-2_channelmaps_mom0_col}.  Contour levels are plotted at $(0.25, 0.5, 1, 2, 4, 8) \times 7.1 {\rm \ Jy \ km \ s^{-1}}$.  To preserve clarity, the moment 1 map was constructed without primary beam correction, by contrast with the moment 0 map.  The cross indicates the position of the continuum emission from the AGN.  The synthesized beam is indicated by the ellipse at the lower left corner.  The large black circle in each panel indicates the primary beam of the SMA antennas.}
\label{CO3-2_mom0_mom1}
\end{figure}
\clearpage

\begin{figure}
\center
\vspace{-4cm}
\epsscale{1.0}
\plotone{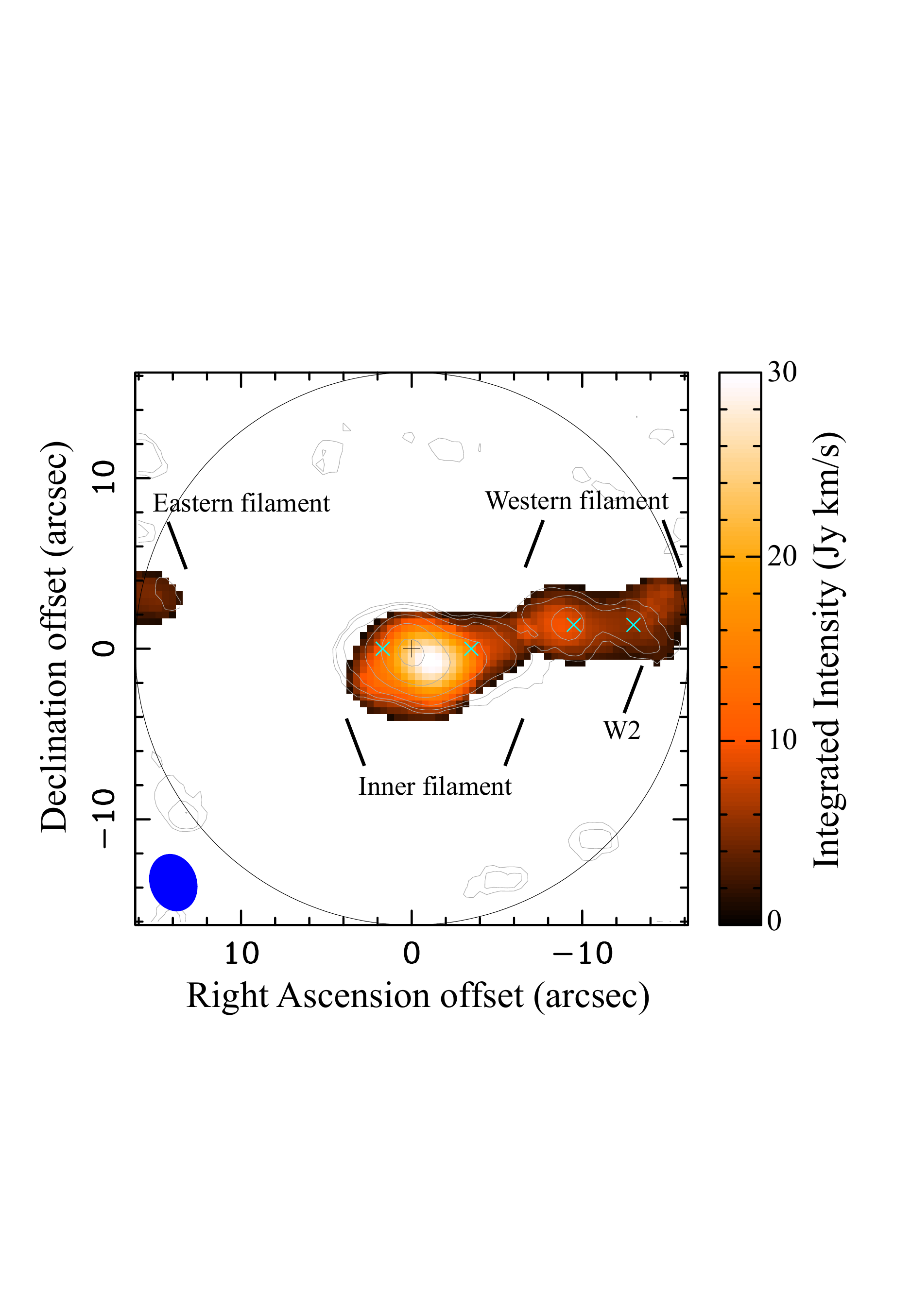}
\vspace{-5.5cm}
\renewcommand{\baselinestretch}{1.0}
\caption{Contours showing integrated intensity in CO(3-2) (same as that shown in colors in Figs.\,\ref{CO3-2_channelmaps_mom0_col} and in contours in \ref{CO3-2_mom0_mom1}) overlaid on colors showing integrated intensity in CO(2-1) \citep[the latter from][]{hol09}.  The filaments are labeled according to the nomenclature introduced by \citet{lim08}.  The $+$ symbol indicates the position of the continuum emission from the AGN.  The four $\times$ symbols, along with the $+$ symbol, indicate the positions at which the spectra shown in Fig.\,\ref{line_profiles} were extracted.}
\label{CO3-2_mom0_CO2-1_mom0}
\end{figure}
\clearpage

\begin{figure}
\center
\vspace{-0.5cm}
\epsscale{1.02}
\plotone{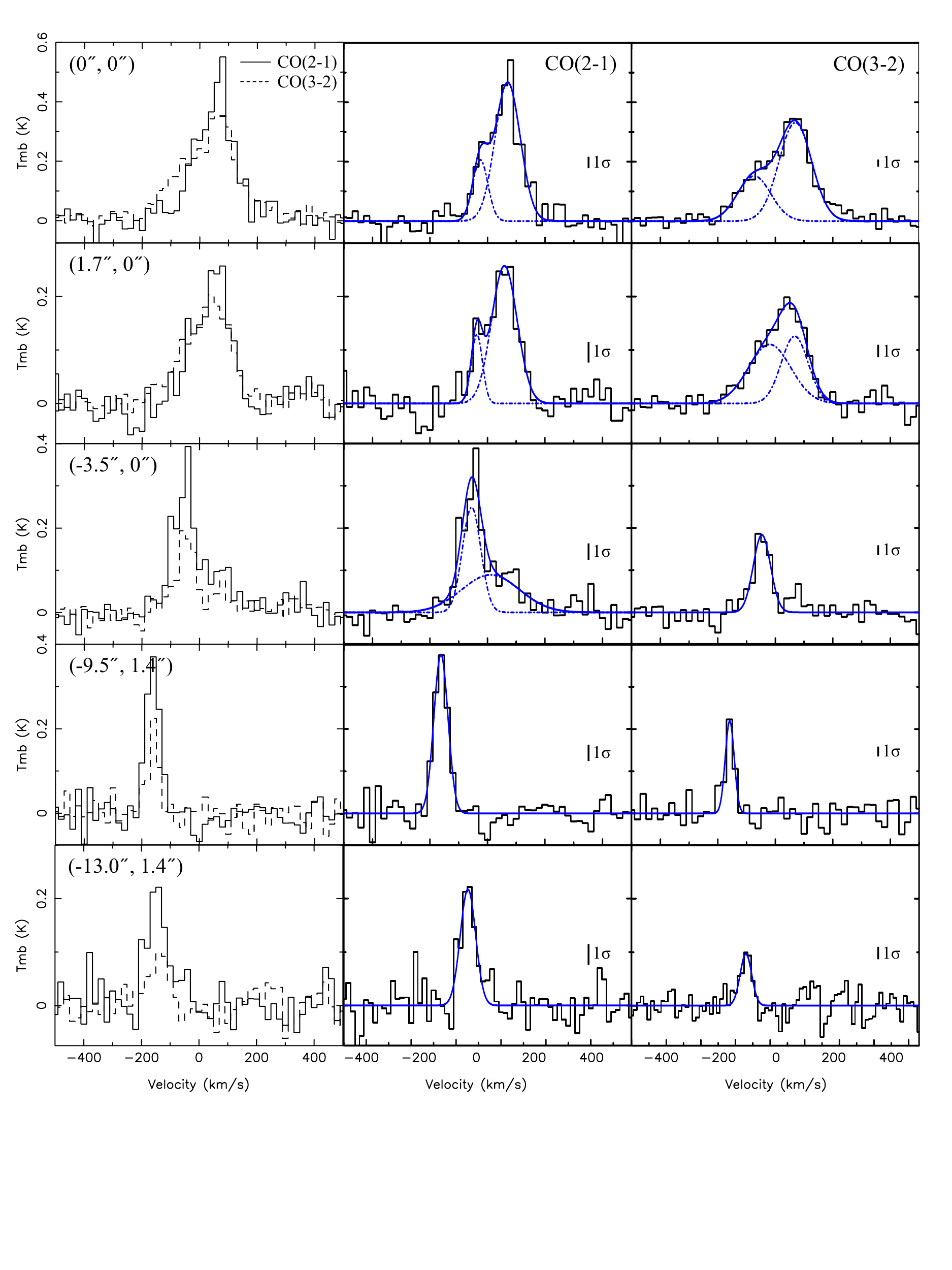}
\vspace{-3.5cm}
\renewcommand{\baselinestretch}{1.0}
\caption{Left column shows line profiles in CO(2-1) (solid line) and CO(3-2) (dashed line) at the positions labeled in each panel as indicated by the $+$ and $\times$ symbols in Fig.\,\ref{CO3-2_mom0_CO2-1_mom0}.  Middle and right columns show the line profiles (black line) in CO(2-1) and CO(3-2), respectively.  Either one or two Gaussian components (blue lines) are fitted to each of these line profiles.  Where two Gaussian components are fitted, the individual components are shown by the dash-dot lines and the sum of the two components by the solid line.  The root-mean-square uncertainty ($\sigma$) of the measurements is indicated in each panel.}
\label{line_profiles}
\end{figure}
\clearpage

\begin{figure}
\center
\vspace{-3cm}
\epsscale{1.06}
\plotone{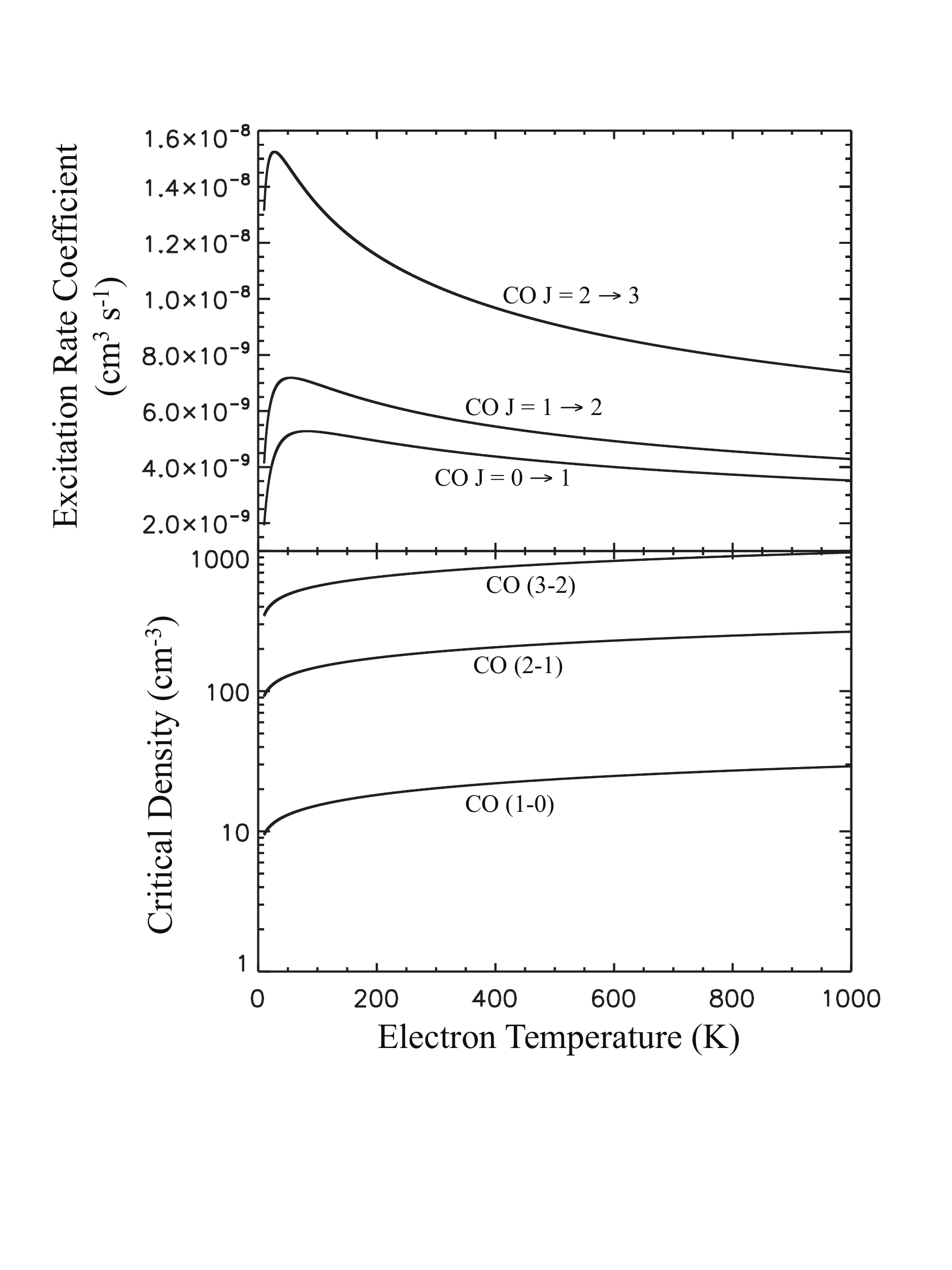}
\vspace{-4.0cm}
\renewcommand{\baselinestretch}{1.0}
\caption{Upper panel shows the electron-impact excitation rate coefficients for CO\,$J = 0 \rightarrow 1$, $1 \rightarrow 2$, and $2 \rightarrow 3$ as a function of electron temperature, computed from Eq.\,(2.21) of \citet{dic75}.  Lower panel shows the corresponding critical density for electron-impact excitation of CO(1-0), CO(2-1), and CO(3-2).}
\label{CO excitation rate}
\end{figure}
\clearpage

\begin{figure}
\center
\vspace{-3cm}
\epsscale{1.06}
\plotone{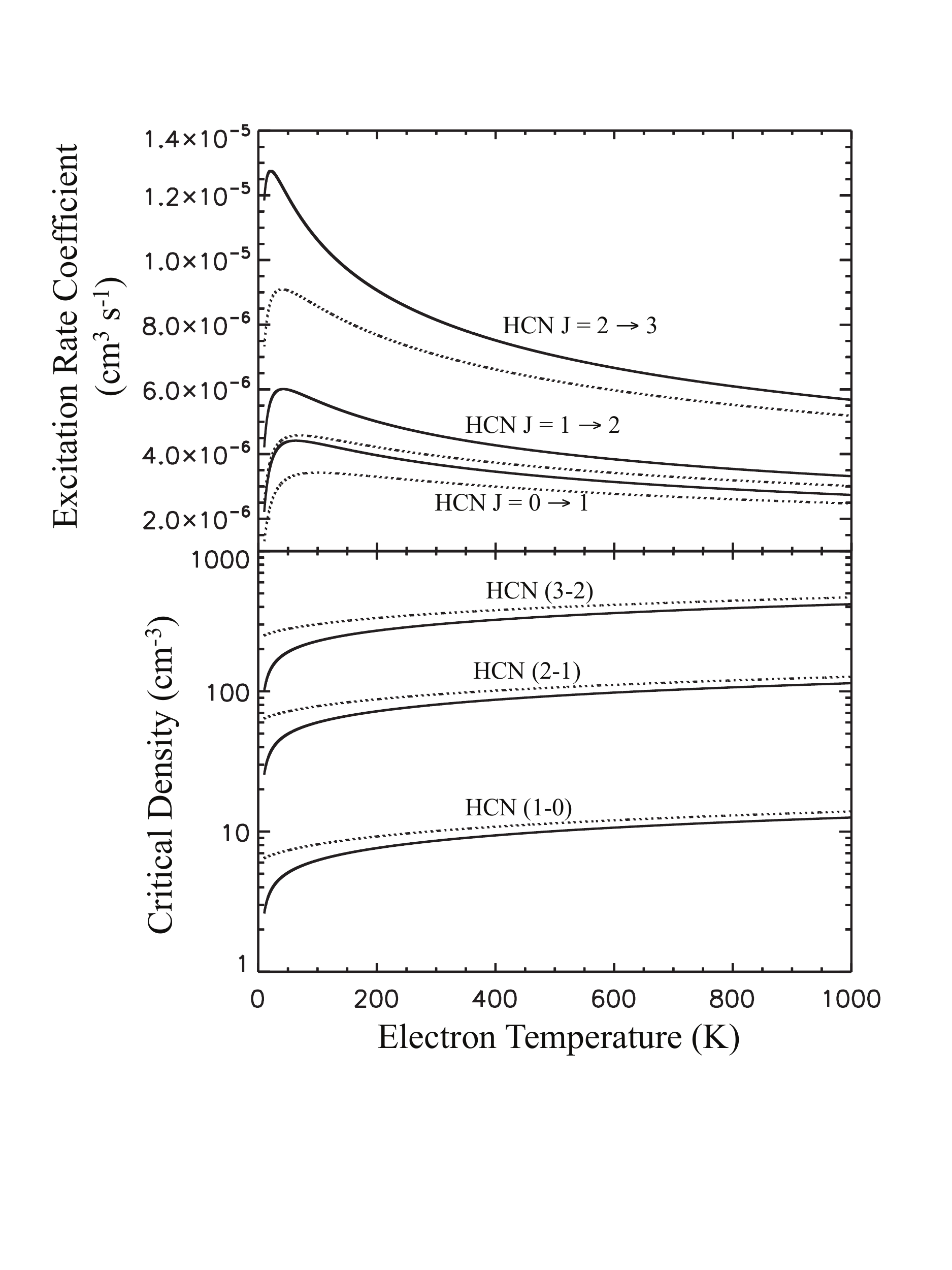}
\vspace{-4.0cm}
\renewcommand{\baselinestretch}{1.0}
\caption{Upper panel shows the electron-impact excitation rate coefficients for HCN\,$J = 0 \rightarrow 1$, $1 \rightarrow 2$, and $2 \rightarrow 3$ as a function of electron temperature.  Lower panel shows the critical density for electron-impact excitation of HCN(1-0), HCN(2-1), and HCN(3-2).  Solid lines were computed from Eq.\,(2.21) of \citet{dic75}, and dashed lines taken from the more recent computations by \citet{fau07a}.}
\label{HCN excitation rate}
\end{figure}
\clearpage

\begin{figure}
\center
\vspace{-3cm}
\epsscale{1.06}
\plotone{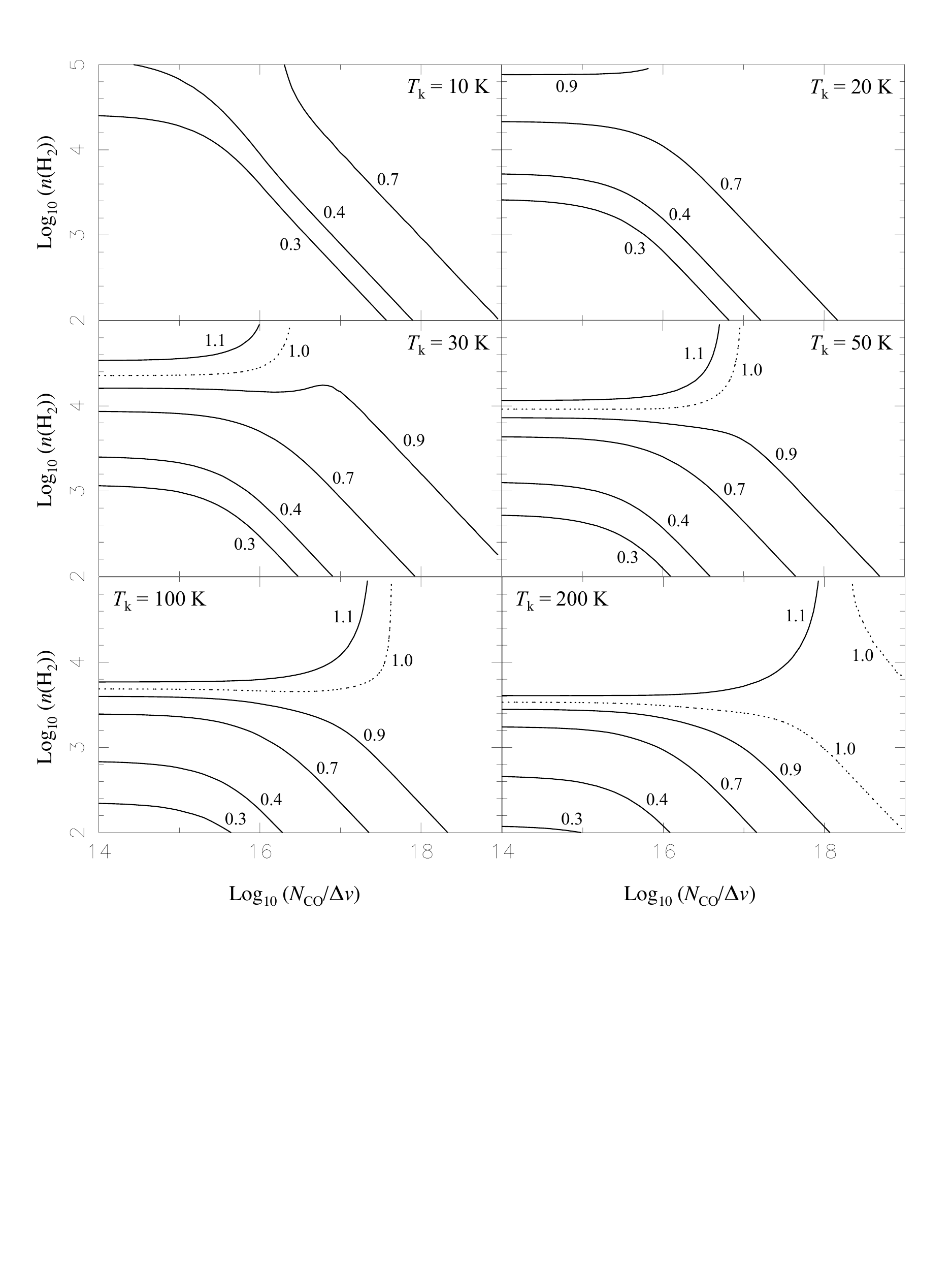}
\vspace{-7.0cm}
\renewcommand{\baselinestretch}{1.0}
\caption{Predicted ratio in brightness temperature of CO(3-2) to CO(2-1), referred to as CO(3-2)/CO(2-1) in the text, as a function of both H$_2$ gas density, $n(\rm H_2)$, and column density in CO per unit velocity, $N_{\rm {CO}}/\Delta v$, at selected gas temperatures, $T_{\rm k}$, as indicated in each panel. }
\label{line_ratio_theory}
\end{figure}
\clearpage

\begin{figure}
\center
\vspace{-3cm}
\epsscale{1.0}
\plotone{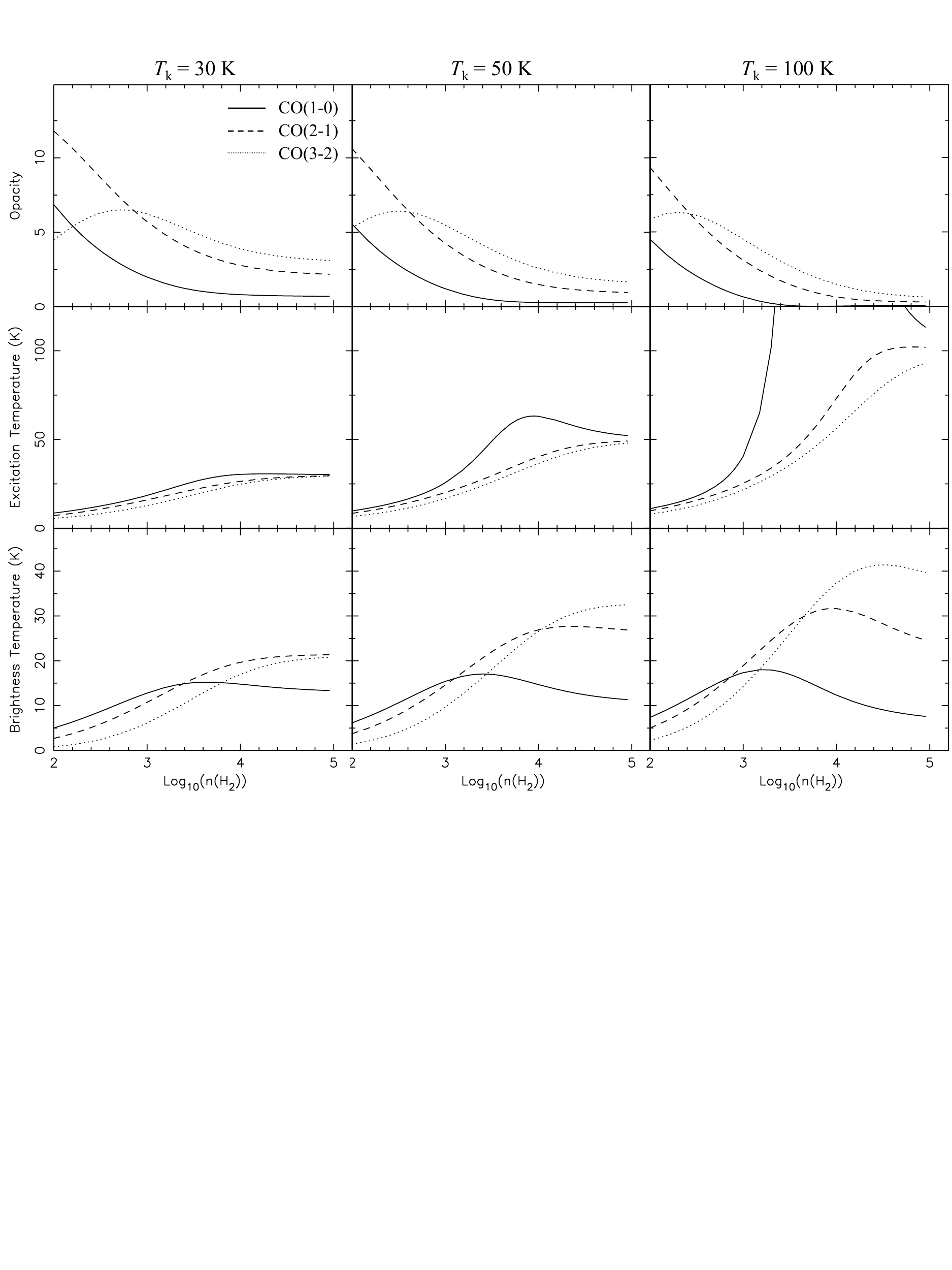}
\vspace{-9.0cm}
\renewcommand{\baselinestretch}{1.0}
\caption{Optical depths (top row), excitation temperatures (middle row), and brightness temperatures (bottom row) of the CO(1-0), CO(2-1), and CO(3-2) lines as a function of H$_2$ gas density, $n(\rm H_2)$, at a column density in CO per unit velocity of $N_{\rm {CO}}/\Delta v = 3 \times 10^{16} {\rm \ cm^{-2} \ km^{-1} \ s}$ and at gas temperatures of $T_{\rm k} = 30$\,K (left column), 50\,K (middle column), and 100\,K (right column).  The CO(1-0) line is suprathermally excited ($T_{\rm ex} > T_{\rm k}$) over the density range $\sim$$10^3$--$10^5 {\rm \ cm^{-3}}$ at all three temperatures.}
\label{CO model 1}
\end{figure}
\clearpage

\begin{figure}
\center
\vspace{-3cm}
\epsscale{1.0}
\plotone{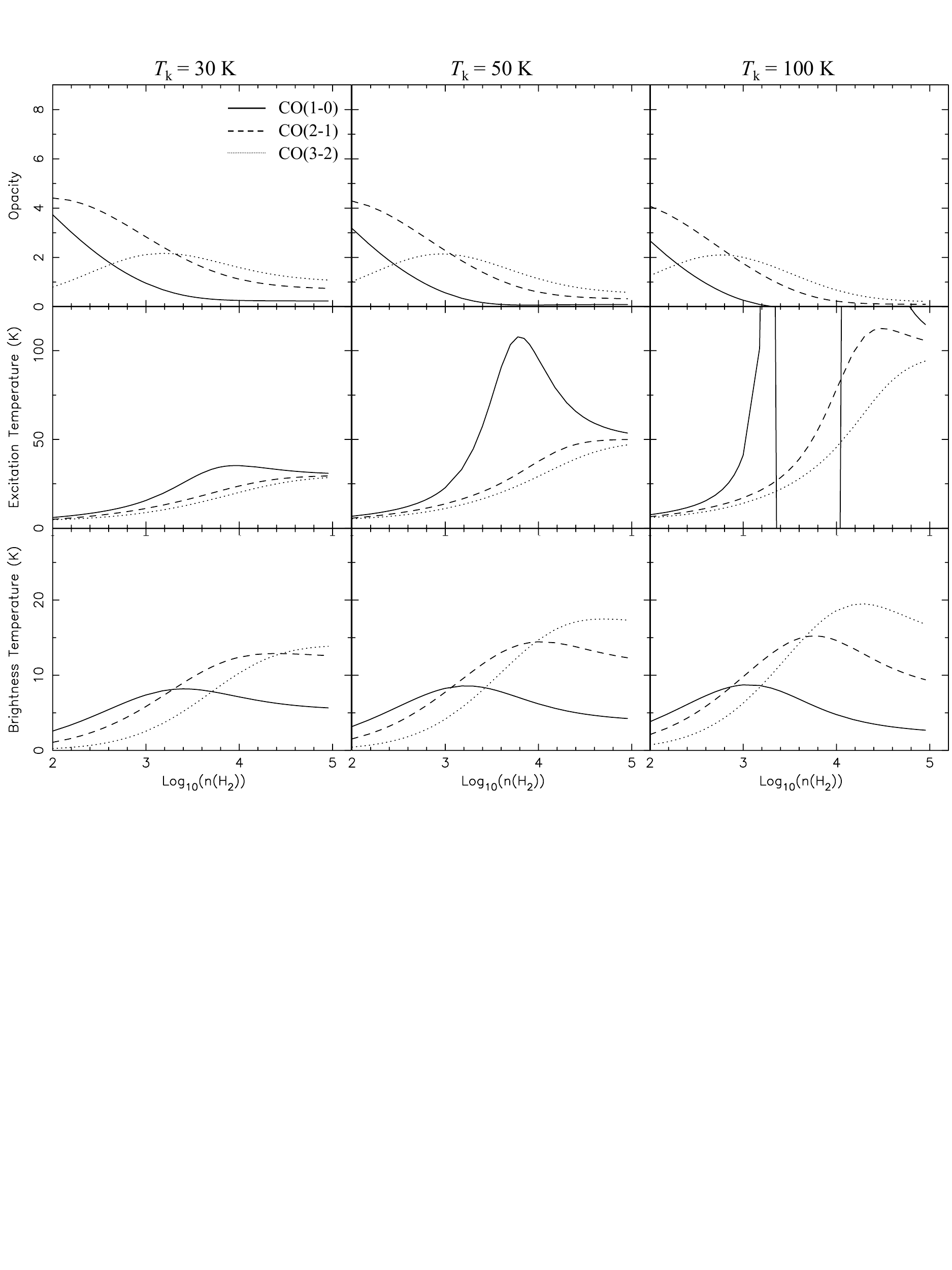}
\vspace{-9.0cm}
\renewcommand{\baselinestretch}{1.0}
\caption{Same as Fig.\,\ref{CO model 1} but at a column density in CO per unit velocity of $N_{\rm {CO}}/\Delta v = 1 \times 10^{16} {\rm \ cm^{-2} \ km^{-1} \ s}$.  The excitation temperature of CO(3-2) becomes negative, indicating a population inversion, over a relatively narrow density range spanning $\sim$$10^{3.4}$--$10^4 {\rm \ cm^{-3}}$.}
\label{CO model 2}
\end{figure}
\clearpage

\begin{figure}
\center
\vspace{-3cm}
\epsscale{1.2}
\plotone{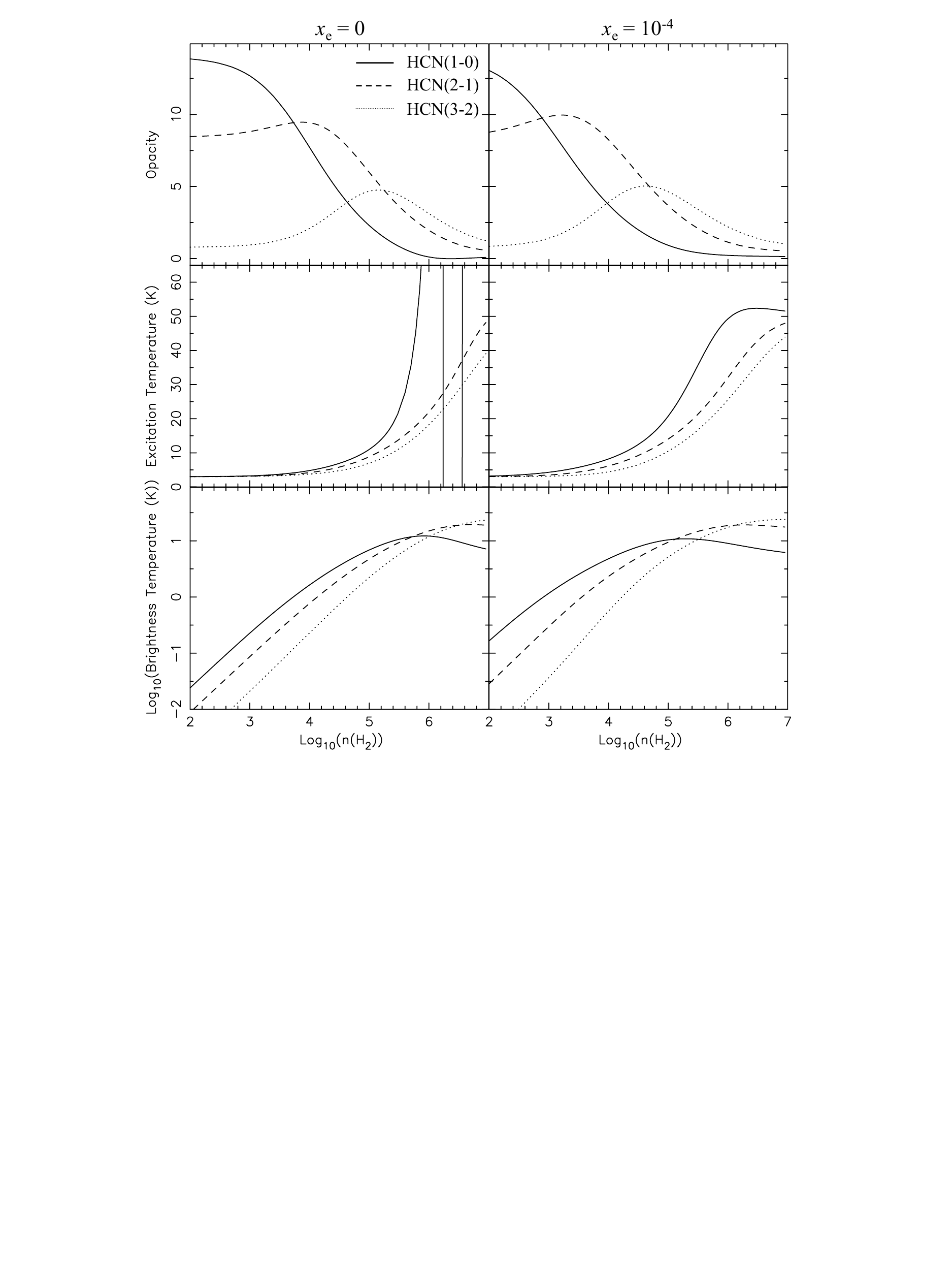}
\vspace{-11.2cm}
\renewcommand{\baselinestretch}{1.0}
\caption{Optical depths (top row), excitation temperatures (middle row), and brightness temperatures (bottom row) of the HCN(1-0), HCN(2-1), and HCN(3-2) lines as a function of H$_2$ gas density, $n(\rm H_2)$, at a column density in HCN per unit velocity of $N_{\rm {CO}}/\Delta v = 3 \times 10^{13} {\rm \ cm^{-2} \ km^{-1} \ s}$ and a gas temperature of $T_{\rm k} = 50$\,K, but for entirely neutral molecular gas (left column) and for molecular gas with an ionization fraction of $x_{\rm e} = 10^{-4}$ (right column).  Note that $x_{\rm e} \equiv n({\rm e}) / 2 \, n({\rm H_2})$, where $n({\rm e})$ is the electron density, as explained in the text.}
\label{HCN model}
\end{figure}
\clearpage

\begin{figure}
\center
\vspace{-3cm}
\epsscale{1.0}
\plotone{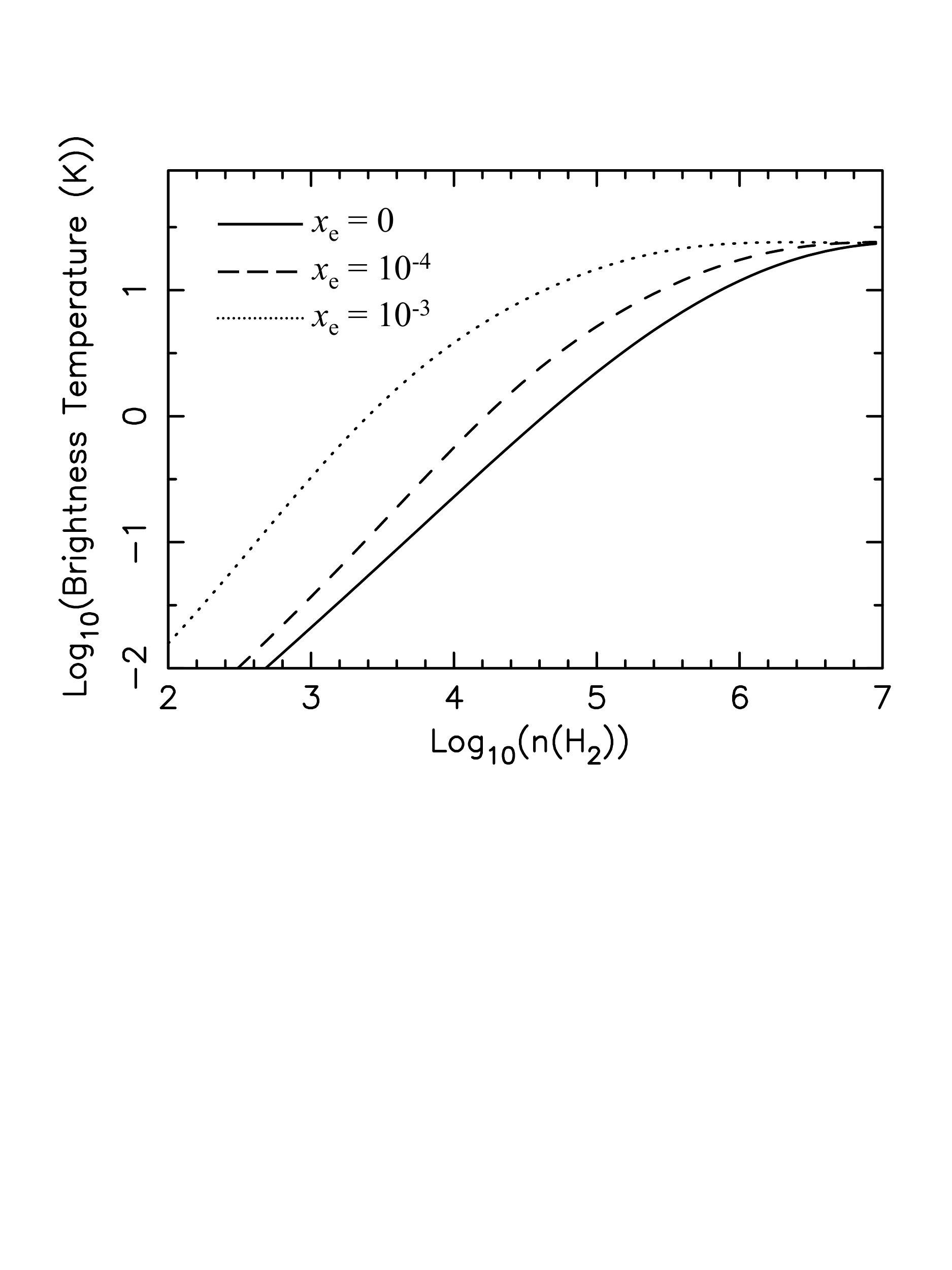}
\vspace{-9.2cm}
\renewcommand{\baselinestretch}{1.0}
\caption{Brightness temperature in HCN(3-2) as a function of H$_2$ gas density, $n(\rm H_2)$, at a column density in HCN per unit velocity of $N_{\rm {HCN}}/\Delta v = 3 \times 10^{13} {\rm \ cm^{-2} \ km^{-1} \ s}$ and a gas temperatures of $T_{\rm k} = 50$\,K (i.e., the same physical parameters as in Fig.\,\ref{HCN model}) for three different ionization fractions of $x_{\rm e} = 0$ (solid line), $x_{\rm e} = 10^{-4}$ (dashed line), and $x_{\rm e} = 10^{-3}$ (dotted line).}
\label{HCN(3-2)}
\end{figure}
\clearpage

\begin{deluxetable}{ccccccc}
\tabletypesize{\small}
\tablecolumns{7}
\tablewidth{0pc}
\tablecaption{Fitted Gaussian Parameters and Line Ratios  \label{line parameters}}
\tablehead{
\colhead{Line} & \colhead{Position$\rm ^a$} & \colhead{Central Velocity$\rm ^b$} &  \colhead{Peak T$_B$} & \colhead{Integrated T$_B$} & \colhead{FWHM} & \colhead{Line Ratio} \\
\colhead{} & \colhead{($\Delta$x\arcsec,\,$\Delta$y\arcsec)} & \colhead{($\rm km \ s^{-1}$)} & \colhead{(K)} & \colhead{($\rm K \ km \ s^{-1}$)} & \colhead{($\rm km \ s^{-1}$)} & \colhead{CO(3-2)/CO(2-1)}}
\startdata
CO(2-1) & \multirow{2}{*}{(0,0)} & 70.0$\rm ^c$ & $0.47$ & $52.5 \pm 3.3$ & $104.8 \pm 8.1$ & \multirow{2}{*}{$0.87 \pm 0.14$} \\
CO(3-2) &                                 & $70.0 \pm 5.2$ & $0.33$ & $45.8 \pm 4.4$ & $130.7 \pm 10.9$ &  \\
\\
CO(2-1) & \multirow{2}{*}{(0,0)} & $-25.0 \pm 4.3$ & $0.21$ & $13.7 \pm 2.4$ & $62.6 \pm 10.1$ & \multirow{2}{*}{---} \\
CO(3-2) &                              & $-71.2 \pm 11.9$ & $0.15$ & $22.2 \pm 4.3$ & $136.8 \pm 19.5$ &  \\
\\
CO(2-1) & \multirow{2}{*}{(0,0)} &     & \multirow{2}{*}{Total:} & $66.2 \pm 5.7$ &                                & \multirow{2}{*}{$0.86 \pm 0.21$} \\
CO(3-2) &                                 &      &                                    & $57.0 \pm 8.7$ &                               &  \\
\\
CO(2-1) & \multirow{2}{*}{(1.7,0)} & $58.8 \pm 6.1$ & $0.26$ & $28.8 \pm 3.3$ & $104.8 \pm 13.9$ & \multirow{2}{*}{$0.49 \pm 0.46$} \\
CO(3-2) &                                 & $67.6 \pm 8.8$ & $0.12$ & $14.1 \pm 11.6$ & $109.8 \pm 24.8$ &  \\
\\
CO(2-1) & \multirow{2}{*}{(1.7,0)} & $-36.7 \pm 6.4$ & $0.13$ & $6.3 \pm 3.3$ & $46.0 \pm 23.8$ & \multirow{2}{*}{---} \\
CO(3-2) &                                 & $-16.6 \pm 43.8$ & $0.11$ & $21.6 \pm 12.1$ & $179.8 \pm 43.3$ &  \\
\\
CO(2-1) & \multirow{2}{*}{(1.7,0)} &     & \multirow{2}{*}{Total:} & $35.1 \pm 6.6$ &                                & \multirow{2}{*}{$1.02 \pm 0.87$} \\
CO(3-2) &                                 &      &                                   & $35.7 \pm 23.7$ &                               &  \\
\\
CO(2-1) & \multirow{2}{*}{(-3.5,0)} & $-46.6 \pm 3.0$ & $0.18$ & $14.3 \pm 1.3$ & $72.9 \pm 8.5$ & \multirow{2}{*}{$0.68 \pm 0.19$} \\
CO(3-2) &                                & $-55.2 \pm 3.3$ & $0.26$ & $21.2 \pm 3.8$ & $77.8 \pm 9.4$ &  \\
\\
CO(2-1) & \multirow{2}{*}{(-3.5,0)} & $22.7 \pm 27.6$ & $0.09$ & $23.5 \pm 5.0$ & $254.1 \pm 45.6$ &  \\
CO(3-2) & & --- &  & & &  \\
\\
CO(2-1) &  \multirow{2}{*}{(-9.5,1.4)}  & $-161.9 \pm 1.7$ & $0.37$ & $22.9 \pm 1.3$ & $57.3 \pm 3.8$ & \multirow{2}{*}{$0.40 \pm 0.08$} \\
CO(3-2) &                                            & $-158.9 \pm 2.4$ & $0.22$ & $9.0 \pm 1.2$ & $38.5 \pm 6.5$ & \\
\\
CO(2-1) & \multirow{2}{*}{(-13,1.4)} & $-146.9 \pm 4.3$ & $0.22$ & $18.5 \pm 2.0$ & $80.4 \pm 10.7$ & \multirow{2}{*}{$0.44 \pm 0.12$} \\
CO(3-2) &                                & $-140.0 \pm 7.1$ & $0.10$ & $8.2 \pm 1.4$ & $79.3 \pm 16.1$ &  \\
\enddata
\tablenotetext{a}{Positions are referenced with respect to the nuclear radio continuum source of \object{NGC 1275}, located at R.\,A.$=$\,03$\rm ^h$\,19$\rm ^m$\,48$\rm ^s$.160 and Dec.$ = +41\degr 30\arcmin 42\farcs103$ (refer to Fig.\,\ref{CO3-2_mom0_CO2-1_mom0}).}
\tablenotetext{b}{Velocity is referenced with respect to the heliocentric velocity of \object{NGC 1275} derived in the optical of $cz = 5264 {\rm \ km \ s^{-1}}$ \citep{huc99}.}
\tablenotetext{c}{Fixed at the central velocity of the corresponding component in CO(3-2).}
\end{deluxetable}
\clearpage

\end{document}